\documentclass[11pt,onecolumn,aps,prc,showpacs,superscriptaddress]{revtex4}

\usepackage{longtable}
\usepackage{latexsym}
\usepackage[dvips]{graphicx}
\usepackage{amssymb}
\usepackage{amsfonts} 
\usepackage{amsmath}

\def\beq{\begin{equation}}
\def\eeq{\end{equation} }
\def\bea{\begin{eqnarray}}
\def\eea{\end{eqnarray}}

\def\bge{\begin{equation}}
\def\ene{\end{equation}}
\def\bgea{\begin{eqnarray}}
\def\enea{\end{eqnarray}}

\usepackage[english]{babel}

\usepackage{indentfirst}

\usepackage{hyperref}
\begin{document}

\title{Electromagnetic form factors of charged and neutral kaons in an extended
vector-meson-dominance model}
\author{S.A. Ivashyn}
\email{ivashin.s@rambler.ru} \affiliation{Kharkov National
University, Kharkov 61077, Ukraine}
\author{A.Yu. Korchin}
\email{korchin@kipt.kharkov.ua} \affiliation{NSC ``Kharkov
Institute of Physics and Technology'', Kharkov 61108, Ukraine}


\begin{abstract}
A model is developed for electromagnetic form factors of the
charged and neutral $K-$mesons. The formalism is based on ChPT
Lagrangians with vector mesons. The form factors, calculated
without fitting parameters, are in a good agreement with
experiment for space-like and time-like photon momenta.
Contribution of the two-kaon channels to the muon anomalous
magnetic moment $a_\mu$ is calculated.
\end{abstract}


\maketitle \frenchspacing \normalsize

\normalsize

\section{Introduction}
\label{sec:introduction}

Hadronic contribution to the vacuum polarization plays an
important role in the test of the Standard Model at the
electroweak precision level. It is the main source of the
theoretical uncertainties in value of the muon anomalous magnetic
moment $a_{\mu}=(g_\mu -2)/2$. The dominant piece of the hadronic
contribution is the $\pi^+\pi^-$ channel, which is expressed in
terms of the pion electromagnetic form factor (FF) $F_\pi (s)$.
This FF has been studied during the last years both experimentally
\cite{Akhmetshin_02} and theoretically (e.g., in hadronic models
\cite{Klingl,Dubinsky_04}). The two-pion channel at low energies
accounts for about 70\% of the leading order (LO) total hadronic
contribution to $a_{\mu}$, at the same time the other hadronic
channels are also important. High precision measurements of
various hadronic channels were performed by CMD-2 and SND in
Novosibirsk, and of $\pi^+ \pi^-$ channel by KLOE in Frascati.

In this paper we concentrate on the electromagnetic (EM) FF's of
the kaons, the charged ones $K^+$, $K^-$, and the neutral ones
$K^0$, $\bar{K}^0$, mainly in the time-like region of the momentum
transfer. Experimental information on these FF's comes from
measurement of the total cross section of the electron-positron
annihilation $e^+e^- \to K^+ K^-$:
\begin{equation}
\label{eq:cross-section} \sigma(e^+e^-\rightarrow K^+ K^-
)=\frac{\pi\alpha^2}{3s} {(1-\frac{4m_{K}^2}{s})^{3/2}}
|F_{K^+}(s)|^2
\end{equation}
and similarly for $e^+ e^- \to K_S K_L$. In
(\ref{eq:cross-section}) \ $s$ is the squared total energy in
center of mass (CM) frame, $\alpha$ is the fine-structure
constant, $m_{K}$ is the $K-$meson mass (the electron mass in
(\ref{eq:cross-section}) is neglected).

New accurate data on the neutral kaon FF's have recently been
obtained in Novosibirsk (CMD-2 collaboration \cite{expAkhmetshin}
and SND collaboration \cite{expAchasov_2005}), and results for the
charged kaon are being analyzed. The kaon FF's are conventionally
studied in terms of the intermediate $J^P =1^{-}$ mesons: \ $\rho
(770) $, $\omega(782)$ and $\phi(1020)$. It turns out that at
energies above $\sqrt{s} \approx 1$ GeV vector resonances
$\rho^\prime = \rho(1450)$, $\omega^\prime = \omega(1420)$ and
$\phi^\prime = \phi(1680)$, which are the low-lying radial
excitations of the $\rho (770) $, $\omega(782)$ and $\phi(1020)$
respectively, also begin to play an important role. The properties
of the latter resonances are not well known. Recently a model has
been worked out \cite{Bruch_04} which accounts for main
vector-meson contributions. The parameters within approach~\cite{Bruch_04}
have been fitted to existing data on pion and kaon FF's.

In the present paper we develop a model for the charged and
neutral kaon EM FF's in both the time-like region (at $\sqrt{s} =
1-2$ GeV) and space-like region, $s \le 0$. It is based on ChPT
with vector mesons \cite{Ecker1,Ecker2}. On the tree-diagrams
level the process $e^+e^- \to K^+ K^-$ (or $e^+ e^- \to K^0
\bar{K}^0$) is described via intermediate $\rho(770) $,
$\omega(782)$ , $\phi(1020)$ states (and possibly including other
$J^P =1^{-}$ resonances), the couplings of which to the photon,
$f_{V} $, are related by the flavor $SU(3)$ symmetry. The
couplings to the kaons, $g_{VK^+K^-}$ and $g_{V K^0 \bar{K}^0}$,
also obey $SU(3)$ symmetry relations. It is also possible to
determine some of these couplings from decay widths, and this way
is preferable as the $SU(3)$ symmetry is not exact. In this model
the photon vector-meson vertices by construction depend on the
photon momentum $q$ and explicitly guarantee gauge invariance of
the theory. These properties are important because they
automatically give the correct normalization of FF's, $ F_{K^+
}(q^2 =0)=1$ and $F_{K^0}(q^2 =0) =0$, naturally account for the
energy dependence of the EM vertices and the vertices vanish in
the case of real-photon vector meson transition.

In the model we include certain loop corrections beyond the tree
level. In particular, there are self-energy polarization operators
that modify vector-meson propagators and loop corrections that
lead to ``dressing'' of the photon-meson vertices. For
construction of vertices in the odd-intrinsic-parity sector, when
necessary, we apply the chiral anomalous Lagrangian of
Wess-Zumino-Witten (WZW) \cite{WZW1,WZW2} for $\gamma \Phi \Phi
\Phi$ interaction and chiral Lagrangians \cite{Ecker3,Prades} for
$V \gamma \Phi$, \ $V \Phi \Phi$ and $V \Phi \Phi \Phi$
interactions. As an option we also use method of
Ref.~\cite{Rudaz1} to constrain parameters of the interaction of
the vector mesons. Parameters of the model are fixed from decay
widths of the vector mesons, and the kaon FF's are calculated
without fitting the parameters.

The plan of the paper is as follows. Sect.~\ref{sec:Formalism}
contains formalism needed to describe the photon, vector-meson and
pseudoscalar-meson interactions. The formalism is based on ChPT
Lagrangians with vector mesons in the even and odd
intrinsic-parity sectors. In sect.~\ref{sec:Form-factors} this
formalism is applied to calculation of the kaon form factors. The
model is extended by using the ``dressed'' propagators and
vertices, and adding the higher resonances. Results of
calculations and discussion are presented in
sect.~\ref{sec:Results}. The $K \bar{K}$--channel contribution to
muon anomalous magnetic moment $a_\mu$ is also calculated. In
sect.~\ref{sec:Conclusions} we draw conclusions. Appendix~A
contains ChPT Lagrangian with vector mesons.

\section{Formalism} \label{sec:Formalism}

\subsection{ChPT Lagrangian}

In the even-intrinsic-parity sector we apply ChPT Lagrangian of
Ecker et al.~\cite{Ecker1,Ecker2} (see Appendix~A for details)
which is an appropriate model in the energy region of interest.
The antisymmetric tensor formulation for the spin-1 fields is used
as it directly fulfills the high-energy constraints of QCD (the
model I of \cite{Ecker2}). The other possibility, vector
formulation, is shown~\cite{Ecker2} to be equivalent to the tensor
formulation to order ${\cal O}(p^4)$ only if additional local
terms of order ${\cal O}(p^4)$ are present in Lagrangian.

Expansion in powers of momenta describes standard EM interaction
of the charged pseudoscalar fields $\Phi$
\begin{eqnarray}
\mathcal{L}_{\gamma \Phi \Phi}^{(1)}& = & \imath e B_\mu
\mathrm{Tr} (Q[\partial_\mu \Phi,\Phi]) = -\imath e B_\mu ( \pi^+
\hat{\partial}^\mu \pi^- + K^+ \hat{\partial}^\mu K^- ),
\label{eq:F1}
\\
\mathcal{L}_{\gamma \gamma \Phi \Phi}^{(2)} &= & -
\frac{e^2}{2}B^\mu B_\mu \mathrm{Tr} \bigl([\Phi,Q]^2 \bigr) = e^2
B^\mu B_\mu (\pi^+\pi^- + K^+K^-), \label{eq:F2}
\end{eqnarray}
where notation $\pi^+ \hat{\partial}^\mu \pi^- \equiv \pi^+
(\partial^\mu \pi^-) - (\partial^\mu \pi^+)  \pi^- $, etc. has
been introduced and $e=\sqrt{4 \pi \alpha} \approx 0.303$ (see
Appendix~A for definition of $\Phi$ and quark charge matrix $Q$).
One also obtains direct photon vector-meson coupling
 \beq
\mathcal{L}_{\gamma V}= e \frac{F_V}{\sqrt{2}} F^{\mu \nu}
\mathrm{Tr}(V_{\mu\nu}Q) =  e F_V  F^{\mu \nu} \bigl(
\frac{1}{2}\rho^0_{\mu\nu} + \frac{1}{6}\omega_{\mu\nu} -
\frac{1}{3\sqrt{2}}\phi_{\mu\nu} \bigr). \label{eq:F3}
 \eeq
This Lagrangian is explicitly gauge invariant and leads to
momentum-dependent vertices for the $ \langle \gamma (\mu) | V(\nu
\rho ) \rangle$ transition, which are proportional to \ $ e F_V
(q^\nu g^{\rho \mu} -q^\rho g^{\mu \nu} )$, where $\mu,\nu,\rho
=0,...,3$ are the Lorentz indices.

The interaction of vector mesons with pseudoscalars is given by
\begin{eqnarray}
\mathcal{L}_{V\Phi\Phi}&=&\imath
\frac{\sqrt{2}G_V}{F_\pi^2} \mathrm{Tr}(V_{\mu\nu}
\partial^\mu\Phi\partial^\nu\Phi) \nonumber \\
&=&\imath \frac{G_V}{F_\pi^2} \big[ \rho^0_{\mu\nu} ( 2
\partial^\mu\pi^+
\partial^\nu \pi^- + \partial^\mu K^+\partial^\nu K^- -
\partial^\mu {K}^0 \partial^\nu \bar{K}^0 )
\nonumber
\\
&&+ \omega_{\mu\nu} \bigl(
\partial^\mu K^+\partial^\nu K^- +
\partial^\mu K^0\partial^\nu\bar{K}^0 )
+ \phi_{\mu\nu} \bigl( -\sqrt{2}
\partial^\mu K^+\partial^\nu K^- - \sqrt{2}
\partial^\mu K^0\partial^\nu \bar{K}^0 ) \big].
\label{eq:F6}
\end{eqnarray}
In deriving (\ref{eq:F6}) the antisymmetry of the vector fields,
$V_{\mu \nu} = - V_{\nu \mu}$, has been used. The pion weak-decay
constant is $F_\pi= 92.4$ MeV.

The coupling constants $F_V$ and $G_V$ can be found from
experimental widths of decays $\Gamma(\rho \to e^+e^-)$ and
$\Gamma(\rho \to \pi\pi)$, respectively. It will be further
convenient to use other constants, $g$ and $f$, related to $F_V$
and $G_V$. Using the data from \cite{pdg} we obtain
\begin{eqnarray}
&&F_V = 156.35 \ \text{MeV}, \;\;\;\;\;\;\;\; G_V = 65.65 \
\text{MeV},
\nonumber \\
&&f  \equiv \frac{m_\rho}{F_V} = 4.966, \;\;\;\;\;\;\;\; g \equiv
\frac{G_V m_\rho }{F_\pi^2} = 5.965. \label{eq:F9}
\end{eqnarray}

In (\ref{eq:F3}) and (\ref{eq:F6}) the exact $SU(3)$ symmetry is
supposed. To have a more flexible model we introduce for each
vector meson independent EM couplings \ $f_V = ( f_\rho, f_\omega,
f_\phi)$ and strong couplings \ $g_{\rho \pi \pi}, \ g_{\rho K^+
K^-}, \ g_{\rho K^0 \bar{K}^0}, \ g_{\omega K^+ K^-}, \ g_{\omega
K^0 \bar{K}^0} $ and $g_{\phi K^+ K^-}, \ g_{\phi K^0 \bar{K}^0}$.
The $SU(3)$ relations for these constants are shown in
Tables~\ref{tab:EM-couplings} and \ref{tab:strong-couplings}. The
couplings $f_V$ can be directly fixed from the decay widths
$\Gamma(V \to l^+l^-)$ (see Table~\ref{tab:EM-couplings}) and
these values will be used in calculations.

\begin{table}
\begin{center}
\begin{tabular}{|c|c|c|c|}
\hline
    & $\rho^0$  & $\omega$ & $\phi$              \\
\hline
$SU(3): \ f_V$  & $f$      & $3f$     & $-\frac{3}{\sqrt{2}} f$ \\
\hline exp.: \ $\Gamma(V \to e^+ e^-)$, keV   & $7.02 \pm 0.11 $
&$0.60 \pm 0.02 $     & $1.27 \pm 0.04 $  \\
$f_V$ & $4.966\pm 0.038 $   & $17.06\pm 0.29 $ & $-13.38\pm 0. 21 $  \\
\hline
exp.: \ $\Gamma(V \to \mu^+ \mu^-)$, keV   & $6.66 \pm 0.2 $  & $0.76 \pm 0.26 $
& $1.21 \pm 0.08 $  \\
$f_V$ & $5.09\pm 0.08 $  & $15.15\pm 2.8 $ & $-13.71\pm 0.45 $ \\
 \hline
\end{tabular}
\end{center}
\caption{Values of the EM coupling constants for $V=\rho^0,
\omega, \phi$.} \label{tab:EM-couplings}
\end{table}

\begin{table}
\begin{center}
\begin{tabular}{|c|c|c|c|}
\hline
 & $\pi^+ \pi^-$ & $K^+ K^-$ & $K^0 \bar{K}^0$ \\
\hline
$\rho^0$ & $g$ & $\frac{1}{2} g$ & $-\frac{1}{2}g $  \\
$\omega$ & -- & $\frac{1}{2} g$ & $\frac{1}{2}g $  \\
$\phi$   & -- & $-\frac{1}{\sqrt{2}} g$ & $-\frac{1}{\sqrt{2}} g $  \\
\hline
\end{tabular}
\end{center}
\caption{$SU(3)$ values of the vector-meson coupling to two
pseudoscalar mesons. } \label{tab:strong-couplings}
\end{table}

The odd-intrinsic-parity interactions are discussed in the
following subsections.

\subsection{Photon - pseudoscalar interaction}
\label{sec:WZW}

The WZW Lagrangian \cite{WZW1,WZW2} involving interaction of
pseudoscalar mesons with external EM field reads
\begin{eqnarray}
\mathcal{L}_{WZW} &=&
\mathcal{L}_{WZW}^{(1)}+\mathcal{L}_{WZW}^{(2)}, \label{eq:WZW}
 \\
 \mathcal{L}_{WZW}^{(1)} &=& -
\frac{e N_c}{48 \pi^2} \epsilon^{\mu\nu\alpha\beta} B_{\mu}
\mathrm{Tr}\Bigl( Q\bigl[ (\partial_\nu U)(\partial_\alpha
U^+)(\partial_\beta U)U^+ - (\partial_\nu U^+)(\partial_\alpha
U)(\partial_\beta U^+)U \bigr] \Bigr), \nonumber
\\
\mathcal{L}_{WZW}^{(2)} &=& \frac{\imath e^2 N_c}{24 \pi^2}
\epsilon^{\mu\nu\alpha\beta} (\partial_\mu B_\nu)B_{\alpha}
\nonumber
\\
&& \times \mathrm{Tr}\Bigl( Q^2(\partial_\beta U)U^+ + Q^2 U^+
(\partial_\beta U) - \frac{1}{2} Q U Q (\partial_\beta U^+)
+\frac{1}{2}Q U^+ Q (\partial_\beta U) \Bigr), \nonumber
\end{eqnarray}
where $N_c =3$ is the number of quark colors, $\epsilon_{0123}= -
\epsilon^{0123}=1$ and $U=\exp (i\sqrt{2}\Phi /F_{\pi })$.

We keep in (\ref{eq:WZW}) the lowest-order EM interaction with
three and one pseudoscalar mesons:
\begin{eqnarray}
\mathcal{L}_{\gamma \Phi \Phi \Phi}&=& - \frac{\imath \sqrt{2}e
N_c }{12 \pi^2 F_\pi^3} \epsilon^{\mu\nu\alpha\beta} B_\mu
\mathrm{Tr}\bigl(Q
\partial_\nu\Phi \partial_\alpha\Phi \partial_\beta\Phi
\bigr)  \nonumber
\\
&=&\frac{\imath e N_c}{12 \pi^2 F_\pi^3}
\epsilon^{\mu\nu\alpha\beta} B_\mu \Bigl(
\partial_\nu \pi^- \partial_\alpha \pi^+ \partial_\beta \pi^0
+
2\, \partial_\nu \pi^0 \partial_\alpha K^- \partial_\beta K^+ +
\partial_\nu \pi^0 \partial_\alpha \bar{K}^0 \partial_\beta K^0
\Bigr), \label{eq:W0}
\\
\mathcal{L}_{\gamma\gamma\Phi}&=& - \frac{\sqrt{2} e^2 N_c}{8
\pi^2 F_\pi} \epsilon^{\mu\nu\alpha\beta} \partial_\mu B_\nu
\partial_\alpha B_\beta \mathrm{Tr}\bigl(Q^2\Phi\bigr). \label{eq:WA}
\end{eqnarray}
The latter, in particular, describes anomalous $\pi^0 \gamma
\gamma$ interaction
\begin{eqnarray}
\mathcal{L}_{\gamma\gamma \pi^0}&=& - \frac{e^2 N_c}{24 \pi^2
F_\pi} \epsilon^{\mu\nu\alpha\beta} \partial_\mu B_\nu
\partial_\alpha B_\beta \pi^0 .
\end{eqnarray}

\subsection{Vector-pseudoscalar-photon interactions}
\label{subsec:Vector_anomalous}

For the odd-intrinsic-parity interactions of vector mesons we use
chiral Lagrangians in vector formulation for spin-1 fields
\cite{Ecker3,Prades}. As shown in \cite{Ecker3} the use of vector
fields for $1^-$ resonances ensures correct behavior of Green
functions to order ${\cal O}(p^6)$. In the alternative tensor
formulation, however, local terms of ${\cal O}(p^6)$ would be
explicitly needed to fulfill high-energy constraints of QCD.
Expanding the corresponding Lagrangian (see Appendix~A) we find
\begin{eqnarray}
\mathcal{L}_{VV\Phi}&=& -\frac{\sqrt{2} \sigma_V
}{F_\pi}\epsilon^{\mu\nu\alpha\beta} \mathrm{Tr}(\partial_\mu
V_\nu \{\Phi, \partial_\alpha V_\beta \}), \label{eq:W-1}
\\
\mathcal{L}_{V\gamma\Phi} &=& - \frac{4 \sqrt{2} e h_V
}{F_\pi}\epsilon^{\mu\nu\alpha\beta}
\partial_\mu B_\nu \mathrm{Tr}(V_\alpha \{ \partial_\beta\Phi, Q \}
).
\label{eq:W1}
\end{eqnarray}
These terms induce the interactions
\begin{eqnarray}
\mathcal{L}_{\omega\rho\pi}&=& - \frac{4 \sigma_V}{F_\pi}
\epsilon^{\mu\nu\alpha\beta}
\partial_\mu \omega_\nu \bigl( \vec{\pi} \partial_\alpha \vec{\rho}_\beta
\bigr), \label{eq:W2}
 \\
\mathcal{L}_{V\gamma\pi^0}&=&  - \frac{4 \sqrt{2} e h_V }{3
F_\pi}\epsilon^{\mu\nu\alpha\beta}
\partial_\mu B_\nu \bigl( \rho^0_\alpha + 3\,
\omega_\alpha  + 3\,\epsilon_{\omega\phi}\phi_\alpha
\bigr)\partial_\beta\pi^0. \label{eq:W3}
\end{eqnarray}
The last term in (\ref{eq:W3}) represents the OZI-forbidden
process, and $\epsilon_{\omega\phi}=0.058$ \cite{Klingl} is the
$\omega$-$\phi$ mixing parameter.  The $V \Phi \Phi \Phi$
interaction is
\begin{eqnarray}
\mathcal{L}_{V \Phi \Phi \Phi}&=& -\frac{ 2 \imath \sqrt{2}
\theta_V}{F_\pi^3}\epsilon^{\mu\nu\alpha\beta} \mathrm{Tr}(V_\mu
\,
\partial_\nu\Phi \, \partial_\alpha\Phi \, \partial_\beta\Phi)
\nonumber
\\
&=& -\frac{2 \imath \sqrt{2} \theta_V
}{F_\pi^3}\epsilon^{\mu\nu\alpha\beta} \Bigl( \rho^0_\mu \,
\partial_\nu\pi^0 \, \partial_\alpha K^+ \, \partial_\beta K^-  +
\rho^0_\mu \, \partial_\nu\pi^0 \, \partial_\alpha K^0 \,
\partial_\beta \bar{K}^0 + 3\,\omega_\mu \, \partial_\nu\pi^0 \,
\partial_\alpha \pi^+ \, \partial_\beta \pi^-
\nonumber
\\
&&+ \omega_\mu \, \partial_\nu\pi^0 \, \partial_\alpha K^+ \,
\partial_\beta K^- - \omega_\mu \, \partial_\nu\pi^0 \,
\partial_\alpha K^0 \, \partial_\beta \bar{K}^0 +
\sqrt{2}\,\omega_\mu \, \partial_\nu\pi^+ \, \partial_\alpha K^0
\, \partial_\beta K^- \nonumber
\\
&& + \sqrt{2}\,\omega_\mu \, \partial_\nu\pi^- \, \partial_\alpha
K^+ \, \partial_\beta \bar{K}^0 + \frac{1}{\sqrt{2}}\,\phi_\mu \,
\partial_\nu\pi^0 \,
\partial_\alpha K^+ \, \partial_\beta K^- +
\frac{1}{\sqrt{2}}\,\phi_\mu \, \partial_\nu\pi^0 \,
\partial_\alpha \bar{K}^0 \, \partial_\beta K^0 \nonumber \\
&& + \phi_\mu \, \partial_\nu\pi^+ \, \partial_\alpha K^0 \,
\partial_\beta K^- + \phi_\mu \, \partial_\nu\pi^- \,
\partial_\alpha K^+ \, \partial_\beta \bar{K}^0 \Bigr) .
\label{eq:W4}                       
\end{eqnarray}

The parameter $h_V$ is fixed to the value $h_V = 0.039$ from the
width of the $\omega \to \gamma \pi^0$ decay. To obtain $\theta_V$
and $\sigma_V$ we use experimental values \cite{pdg} for the
three-pion decay widths of $\omega$ and $\phi$ mesons
\cite{SchechterMeissner}. It is usually supposed that the
$\phi-$meson decays into the three pions via the $\omega$-$\phi$
mixing (for other options see \cite{Achasov0305049,
NNAchasovPLB233, NNAchasovPRD52}), and that the amplitudes for the
direct decay, $\phi \to \pi \pi \pi$, and decay through the
intermediate state, $ \phi \to \rho\pi \to \pi \pi \pi$, sum up
incoherently. According to \cite{Aloisio0303016, Achasov9907026,
Achasov0309055} the direct $\omega\to \pi \pi \pi$ (and $\phi \to
\pi \pi \pi$) decay is suppressed with respect to the process
$\omega\to \rho\pi \to \pi \pi \pi$ (and $\phi\to \rho\pi \to \pi
\pi \pi$). In view of these constraints we obtain $\sigma_V =
0.33$ and $\theta_V = 0.0011$.

There is an interesting approach \cite{Rudaz1} which generalizes
the WZW anomaly term \cite{WZW1,WZW2} for the case of vector and
axial-vector mesons. The authors of \cite{Rudaz1} applied
Bardeen's form of anomaly to keep the vector current anomaly free
and have anomaly for the axial-vector current. The $V \Phi \Phi
\Phi$ and $V V \Phi$ pieces of Lagrangian in this approach have
the form
\begin{eqnarray}
\mathcal{L}_{V \Phi \Phi \Phi}&=& -\frac{\imath g}{4\pi^2 F_\pi^3}
\epsilon^{\mu\nu\alpha\beta} \mathrm{Tr}(V_\mu \,
\partial_\nu\Phi \, \partial_\alpha\Phi \, \partial_\beta\Phi),
\label{eq:W100}
 \\
\mathcal{L}_{V V\Phi}&=& -\frac{3 g^2}{16 \sqrt{2}\pi^2 F_\pi}
\epsilon^{\mu\nu\alpha\beta} \mathrm{Tr}(\partial_\mu V_\nu \{
\Phi, \partial_\alpha V_\beta \}).
 \label{eq:W101}
\end{eqnarray}
The EM field can be included, based on $U(1)$ gauge invariance and
vector-meson dominance, by the substitution
\begin{equation}
V_\mu \to V_\mu + \frac{\sqrt{2}e}{g}Q B_\mu
 \label{eq:min-subst}
\end{equation}
in the covariant derivative \cite{Rudaz1} acting on $U$. By
substituting (\ref{eq:min-subst}) in (\ref{eq:W101}) one obtains
$V \gamma \Phi$ interaction
\begin{equation}
\mathcal{L}_{V\gamma\Phi}= -\frac{3e g}{8 \pi^2
F_\pi}\epsilon^{\mu\nu\alpha\beta}
\partial_\mu B_\nu \mathrm{Tr}(V_\alpha \{ \partial_\beta\Phi, Q \})
\label{eq:gamma-V-phi_id}          
\end{equation}
and $\gamma \gamma \Phi$ Lagrangian (\ref{eq:WA}). Further,
substitution of (\ref{eq:min-subst}) in (\ref{eq:W100}) leads to
$\gamma \Phi \Phi \Phi$ interaction (\ref{eq:W0}).

Eqs.(\ref{eq:W100}), (\ref{eq:W101}) and (\ref{eq:gamma-V-phi_id})
can be used to estimate parameters $h_V, \theta_V $ and $
\sigma_V$ in (\ref{eq:W-1}), (\ref{eq:W1}) and (\ref{eq:W4}) (we
call the corresponding values ``ideal'' ones). These values are
shown in the 2nd line of Table~\ref{tab:PhRel} and the values
determined from experiment are listed in the 3rd line. It is seen
that the parameters $h_V$ and $ \sigma_V$ agree well, while the
experiment favors very small value of the constant $\theta_V$ for
the $V \Phi \Phi \Phi$ vertex.

In the last line of Table~\ref{tab:PhRel} we present results for
these couplings in an extended Nambu-Jona-Lasinio model of QCD
\cite{Prades}. For a particular value of axial constant $g_A=1$,
and $f^\prime \approx 5.9$ from \cite{Prades}, the obtained
couplings agree with the ``ideal'' values.

\begin{table}
\begin{center}
\begin{tabular}{|c|c|c|c|}
\hline
Parameters & $h_V$ & $\theta_V$ & $\sigma_V$   \\
\hline ``ideal'' values  & $\frac{3 g}{32 \sqrt{2}\pi^2}=0.040$
& $\frac{g}{8 \sqrt{2} \pi^2}=0.054$ & $\frac{3 g^2}{32 \pi^2}= 0.34$ \\
\hline
fixed from experiment & $0.039$ & $0.0011$ & $0.33$\\
\hline values in model \cite{Prades} for $g_A =1$ & $\frac{N_c
f'}{32 \sqrt{2} \pi^2}= 0.040$ & $\frac{N_c f'}{24\sqrt{2}
\pi^2}=0.053$ & $\frac{N_c f'^2}{32 \pi^2 }=0.33$ \\
\hline
\end{tabular}
\end{center}
\caption{Values of parameters in Lagrangians (\ref{eq:W-1}),
(\ref{eq:W1}) and (\ref{eq:W4}). \ $g=5.965$ is fixed from $\rho
\to \pi \pi$ decay, and $f' = 1/0.17 \approx 5.9$ \cite{Prades}.}
\label{tab:PhRel}
\end{table}

\section{Kaon electromagnetic form factors}
\label{sec:Form-factors}

The kaon EM FF's can be defined in terms of the quark EM current
(see, e.g., \cite{Icikson})
 \bea
j^\mu_{em}(x) &=& \frac{2}{3} \bar{u}(x) \gamma^\mu
u(x)-\frac{1}{3} \bar{d}(x) \gamma^\mu d(x) -\frac{1}{3}
\bar{s}(x) \gamma^\mu s(x ) \nonumber \\
&=& V^\mu_3 (x) + \frac{1}{\sqrt{3}} V^\mu_8 (x),
 \eea
where the isovector and isoscalar components are respectively
\bea
V^\mu_{3}(x) &=& \bar{q}\gamma^\mu \frac{\lambda_3}{2}q =
\frac{1}{2}(\bar{u}\gamma^\mu u - \bar{d}\gamma^\mu d),  \nonumber \\
\frac{1}{\sqrt{3}} V^\mu_8 (x) &=& \frac{1}{\sqrt{3}}
\bar{q}\gamma^\mu \frac{\lambda_8}{2}q = \frac{1}{6}
(\bar{u}\gamma^\mu u + \bar{d}\gamma^\mu d) - \frac{1}{3}
\bar{s}\gamma^\mu s. \eea
 Then the FF's are defined through the matrix elements
 \bea
\langle K^{+}(p_1) K^{-} (p_2 ) |j^\mu_{em} (0)| 0 \rangle
& \equiv & (p_1 - p_2 )^\mu F_{K^+} (q^2), \nonumber \\
\langle K^{0}(p_1) \bar{K}^{0} (p_2 ) |j^\mu_{em} (0)| 0 \rangle &
\equiv & (p_1 - p_2 )^\mu F_{K^0} (q^2),
 \eea
where $p_1, p_2$ are the kaon momenta, $q = p_1+ p_2$ and $q^2
\equiv s$. These FF's, being defined in the time-like region $s
\ge 4 m_K^2$, due to the analyticity describe also the space-like
region $s < 0$ corresponding to elastic electron scattering on the
kaon.

The direct $\gamma V$ coupling in (\ref{eq:F3}) leads to the
following form for FF's:
 \beq
F_{K^+} (s) = 1- \sum_{V =\rho,\omega,\phi}\frac{g_{VK^+
K^-}}{f_{V}(s)}A_V(s), \quad \quad F_{K^0} (s) = -
\sum_{V=\rho,\omega,\phi} \frac{g_{VK^0 \bar{K}^0}}{f_{V}(s)}
A_V(s), \label{eq:FFs}
 \eeq
 \beq A_V(s) \equiv \frac{s}{s-m_V^2 - \Pi_V(s)},
 \label{eq:A_V}
 \eeq
where $\Pi_V(s)$ is the self-energy operator for the vector meson
$V=\rho,\omega,\phi$.

It is seen that due to gauge invariance of the photon vector-meson
interaction the coupling is proportional to photon invariant mass
$s$ and vanishes for real photons. This automatically leads to the
correct normalization conditions
 \beq
F_{K^+ }(0)=1, \quad \quad \quad F_{K^0 }(0) =0.
\label{eq:normalization}
 \eeq
Thus there is no need to impose constraints on coefficients in
(\ref{eq:FFs}) in order to obey (\ref{eq:normalization}). An
additional energy dependence of the couplings $f_{V}(s)$ arises
due to higher-order corrections, as described in
subsect.~\ref{subsec:EM-vertex}.

Using for the strong couplings the $SU(3)$ relations in
Table~\ref{tab:strong-couplings} we can write
 \bea
F_{K^+}(s) &=& 1 - \frac{g}{2f_\rho(s)} A_{\rho^0}(s) -
  \frac{g}{2f_\omega(s)} A_\omega (s) +
\frac{g}{\sqrt{2} f_\phi(s)} A_\phi (s),
\label{eq:A1}
\\
F_{K^0}(s) &=& \frac{g}{2 f_\rho(s)} A_{\rho^0}(s) -
  \frac{g}{2 f_\omega(s)} A_\omega (s) +
\frac{g}{\sqrt{2} f_\phi(s)} A_\phi (s).
\label{eq:A2}
 \eea

\subsection{Self-energy operators}
\label{sec:self-energy}

\begin{figure}[htbp]
    \centering
\begin{tabular}{p{3cm}p{3cm}p{3cm}}
\parbox{30mm}{
\includegraphics[width=30mm]{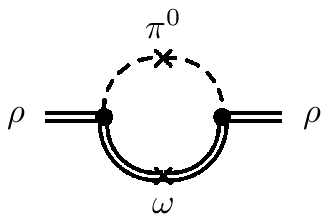}
}&
\parbox{30mm}{
\includegraphics[width=30mm]{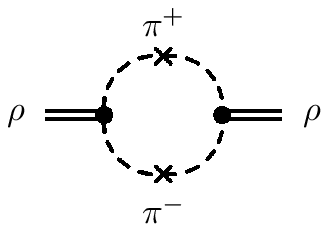}
} &
\parbox{30mm}{
\includegraphics[width=30mm]{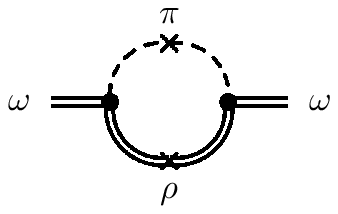}
}
\\
\parbox{30mm}{
\includegraphics[width=30mm]{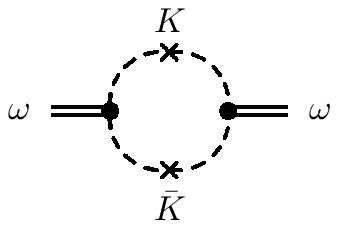}
} &
\parbox{30mm}{
\includegraphics[width=30mm]{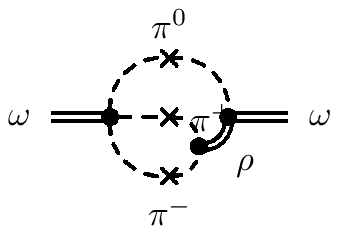}
} &
\parbox{30mm}{
\includegraphics[width=30mm]{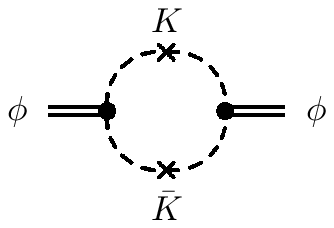}
}
\end{tabular}
    \caption{Loops included in self-energy of vector mesons.
    Crosses indicate imaginary part of the diagrams.}
    \label{fig:loops_se}
\end{figure}

In general the dressed propagator of vector particles includes the
self-energy operators $\Pi_V(s)$. The latter should account for
all intermediate states, such as
$\pi^+\pi^-$, \
$\omega\pi^0$, \ $K \bar{K}$, \
 $\omega\pi^0 \to \pi^0 K^+ K^- $ for $\rho$ meson,
$K \bar{K}$, \ $\rho \pi$, \ $\pi^0\pi^+\pi^-$, \ $\rho \pi \to
\pi^0\pi^+\pi^- $, \ $\rho \pi \to \pi^0 K \bar{K}$ \ for $\omega$
meson, $K \bar{K}$, \ $\rho \pi $, \ $\pi^0 K \bar{K}$, \ $\rho
\pi \to \pi^0 K \bar{K}$, \ $\pi^0\pi^+\pi^- $, \ $\rho \pi \to
\pi^0\pi^+\pi^-$ for $\phi$ meson (including those with
$\omega-\phi$ mixing), and possibly others. As our calculations
indicates, in the region of interest the most important
contributions to $\Pi_V(s)$ consist of the diagrams shown in
Fig.~\ref{fig:loops_se}. These diagrams are the dominant ones in
the energy interval $\sqrt{s}$ about few GeV's, they contribute to
the self-energy for $\rho$, $\omega$ and $\phi$ mesons about 86\%,
95\% and 95\% respectively. We take into consideration that there
is experimental evidence \cite{Aloisio0303016, Achasov9907026,
Achasov0309055} of a relatively small direct couplings $\omega,
\phi \to \pi \pi \pi$.

These diagrams can be collected in the self-energy operators as
follows
\begin{eqnarray}
\Pi_\rho &=& \Pi_{\rho(\pi^0\omega)\rho} + \Pi_{\rho(\pi\pi)\rho},
\\
\Pi_\omega & =& \Pi_{\omega(\pi^0\rho)\omega} +
\Pi_{\omega(KK)\omega} + 2 \Pi_{\omega(3\pi,\pi\rho)\omega},
\\
\Pi_\phi &=& \Pi_{\phi(KK)\phi},
\end{eqnarray}
where the subscript notation is explained in
Fig.~\ref{fig:loops_se}.

In the following we include only the imaginary parts of the loop
contributions. These will be the dominant contributions giving
rise to the energy-dependent widths
 \beq
 \Gamma_V (s) = - m_V^{-1}\ {\rm Im}\, \Pi_V (s).
 \eeq
The real part of the self-energy, ${\rm Re}\, \Pi_V(s)$, usually
can be taken into account by an appropriate renormalization of the
coupling constants and masses. An approximation, consisting in
neglecting the real part of the loop contributions compared to the
imaginary part, is often used in scattering theory and is known as
$K-$matrix approach. For an example of successful application of
the coupled-channel $K-$matrix approach to reactions on the
nucleon see Ref.~\cite{Korchin_96}.

Applying the Cutkosky rules (\cite{Icikson}, Ch.6.3) to the
diagrams shown in Fig.~\ref{fig:loops_se} we find the imaginary
parts: \bea
&& \label{se1}
{\rm Im }
\Pi_{\rho(\pi^0\omega)\rho}(s) =
\frac{-\sqrt{s}}{12\pi}\frac{16\, \sigma^2_V}{F_\pi^2}
(\frac{(s+m_\pi^2-m_\omega^2)^2}{4 s} - m_\pi^2)^{3/2}
\theta(s-(m_\pi+m_\omega)^2),
\label{se2}
\\
&&{\rm Im } \Pi_{\rho(\pi\pi)\rho}(s) = \frac{-g^2
}{48\pi\sqrt{s}} (s-4 m_\pi^2)^{3/2} \theta(s-4 m_\pi^2),
\nonumber
\\
&&{\rm Im } \Pi_{\omega(\pi^0\rho)\omega}(s) =
\frac{-\sqrt{s}}{12\pi}\frac{16 \,\sigma_V^2}{F_\pi^2}
(\frac{(s+m_\pi^2-m_\rho^2)^2}{4 s} - m_\pi^2)^{3/2}
\theta(s-(m_\pi+m_\rho)^2), \nonumber
\\
&&{\rm Im } \Pi_{\omega(KK)\omega}(s) =
\frac{-\frac{1}{2}g^2}{48\pi\sqrt{s}}
 (s-4 m_K^2)^{3/2} \theta(s-4 m_K^2), \nonumber
\\
&&{\rm Im } \Pi_{\phi(KK)\phi}(s) = \frac{-g^2} {48\pi\sqrt{s}}
(s-4 m_K^2)^{3/2} \theta(s-4 m_K^2), \nonumber
\\
&&{\rm Im } \Pi_{\omega(3\pi,\pi\rho)\omega}(s) = \frac{s}{2^8
\pi^3} \int dE_+dE_-\; 4\,\Bigl((\vec{p}_-\vec{p}_+)^2 -
\vec{p}_-{}^2\vec{p}_+{}^2\Bigr) \nonumber \\
&& \times \Bigl[ -\frac{g}{3} \frac{24 \sqrt{2}\,h_V\,\sigma_V}{2
F_\pi^4} {\rm Im } \frac{1}{p_\rho^2 - m_\rho^2 - \imath {\rm Im }
\Pi_\rho}\Bigr], \nonumber
\\
&& 4\,\Bigl((\vec{p}_-\vec{p}_+)^2 -
\vec{p}_-{}^2\vec{p}_+{}^2\Bigr)= [(\sqrt{s}- E_+ - E_-)^2 +
m_\pi^2  - E_-^2 - E_+^2]^2 \nonumber \\
&& \;\;\; - 4(E_+^2 - m_{\pi}^2)(E_-^2 - m_{\pi}^2), \nonumber
\eea
where $\theta(x)=1$ for $x>0$ and $\theta(x)=0$ otherwise,
$E_{\pm} = \sqrt{\vec{p}_{\pm}{}^2 + m_{\pi}^2}$, \ and region of
integration is restricted by
\begin{equation*}
\left\{
\begin{aligned}
m_{\pi} \leq E_{\pm} \leq \sqrt{s} - 2 m_{\pi},
\\
E_+ + E_- \leq \sqrt{s} - m_{\pi},
\\
\vec{p}_+{}^2 \vec{p}_-{}^2 - (\vec{p}_+{}\vec{p}_-{})^2 > 0.
\end{aligned}
\right.
\end{equation*}
Choosing $p_\rho^2 = p_{\rho^0}^2,\,p_{\rho^+}^2,\,p_{\rho^-}^2$
allows one to include all three possible charge states of the most
complicated diagram for $\omega$ meson in Fig.~\ref{fig:loops_se}:
\bea p_{\rho^0}^2 &=& - s +m_\pi^2 + 2 \sqrt{s} (E_+ + E_-),
\nonumber
\\
p_{\rho^\pm}^2 &=&  s +m_\pi^2 - 2 \sqrt{s}E_{\mp} \ . \nonumber
\eea

In order to restrict the fast growth of the partial widths with
$s$ we introduce cut-off FF's for the vertices describing
vector-meson decay into two pseudoscalars and pseudoscalar-vector.
For the particular form of these FF we
choose~\cite{Achasov0305049}
\begin{eqnarray}
C_{VVP}(s,r_0) &=& \frac{1+ (r_0 m_V)^2}{1 + (r_0 \sqrt{s})^2},
\label{eq:K1}
\\
C_{VPP}(s,r_0) &=& \sqrt{\frac{1+(r_0 q_P(m_V^2))^2}{1+ (r_0
q_P(s))^2}}, \label{eq:K2}
\\
q_P (s) &\equiv& \sqrt{s-4 m_P^2}, \quad \quad \quad r_0 = (2.5
\pm 1.1 \pm 0.5) \; \text{GeV}^{-1}. \nonumber
\end{eqnarray}
Here $r_0$ stands for vector-meson effective radius which is taken
the same for all decays. With this FF \ $\Gamma(s) \to const$ as
$s \to \infty$. All the above expressions for the self-energy
operators are multiplied by the corresponding FF squared, \ i.e.
${\rm Im \,}\Pi_{\cdots}(s) \Rightarrow {\rm Im\,}\Pi_{\cdots}(s)
C^2_{\cdots}(s,r_0)$.

\subsection{Electromagnetic vertex modification}
\label{subsec:EM-vertex}

In this subsection we focus on the photon vector-meson vertices,
which follow from (\ref{eq:F1}), (\ref{eq:W0}), (\ref{eq:W2}) and
(\ref{eq:W4}). As mentioned in subsection \ref{sec:self-energy}
the EM couplings $f_V (s)$ acquire energy dependence due to loop
corrections in the present model.

\begin{figure}[htbp]
    \centering
\begin{tabular}{p{3cm}p{3cm}p{3cm}}
\parbox{30mm}{
\includegraphics[width=30mm]{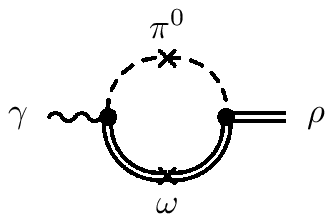}
}&
\parbox{30mm}{
\includegraphics[width=30mm]{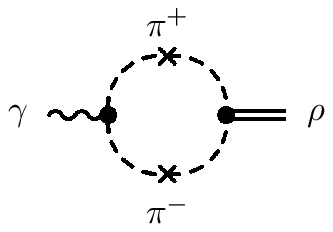}
} &
\parbox{30mm}{
\includegraphics[width=30mm]{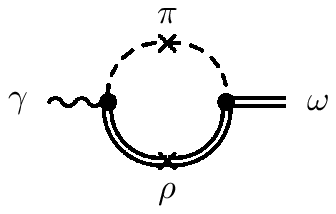}
}
\\
\parbox{30mm}{
\includegraphics[width=30mm]{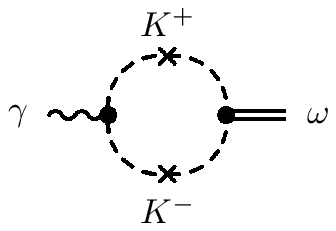}
} &
\parbox{30mm}{
\includegraphics[width=30mm]{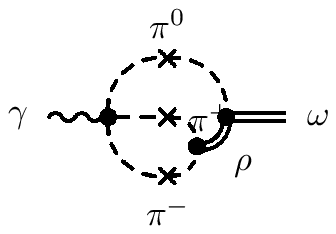}
} &
\parbox{30mm}{
\includegraphics[width=30mm]{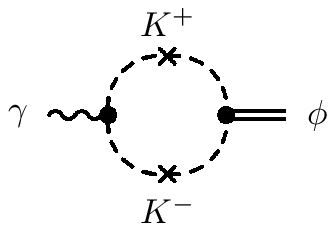}
}
\end{tabular}
    \caption{Loops for EM vertex modification.}
    \label{fig:loops_vertex}
\end{figure}

To be consistent with the approximation for the self-energy we
include only the imaginary part of the loop contributions to the
EM vertex functions which are shown in
Fig.~\ref{fig:loops_vertex}.

In numerical calculation the following formulae are used:
\begin{eqnarray}
\label{eq:E1} {\rm Im }\, \Pi_{\gamma(\pi^0\omega)\rho}(s) =&
\frac{\sqrt{2}e\,h_V}{\sigma_V} {\rm Im }\, \Pi_{\rho(\pi^0\omega)\rho}(s),
\nonumber
\\
{\rm Im }\, \Pi_{\gamma(\pi\pi)\rho}(s) =&
\frac{e}{g} {\rm Im }\, \Pi_{\rho(\pi\pi)\rho}(s),
\nonumber
\\
{\rm Im }\, \Pi_{\gamma(\pi^0\rho)\omega}(s) =& \frac{\sqrt{2}e\,h_V }{3
\sigma_V} {\rm Im }\, \Pi_{\omega(\pi^0\rho)\omega}(s), \nonumber
\\
{\rm Im }\, \Pi_{\gamma(KK)\omega}(s) =& \frac{e}{g} {\rm Im }\,
\Pi_{\omega(KK)\omega}(s), \nonumber
\\
{\rm Im }\, \Pi_{\gamma(KK)\phi}(s) =& -\frac{e}{\sqrt{2} g}
{\rm Im }\, \Pi_{\phi(KK)\phi}(s), \nonumber
\\
{\rm Im }\, \Pi_{\gamma(3\pi,\pi\rho)\omega}(s) = & \frac{e}{24 \sqrt{2}
\pi^2 \, \theta_V} {\rm Im }\, \Pi_{\omega(3\pi,\pi\rho)\omega}(s).
\label{eq:E2}
\end{eqnarray}
These expressions should also be multiplied by cut-off FF $C_{V
\gamma P}(s,r_0)$,  which is chosen equal to $C_{VVP}(s,r_0)$ in
(\ref{eq:K1}).

\begin{figure}[htbp]
\begin{center}
\includegraphics[width=140mm]{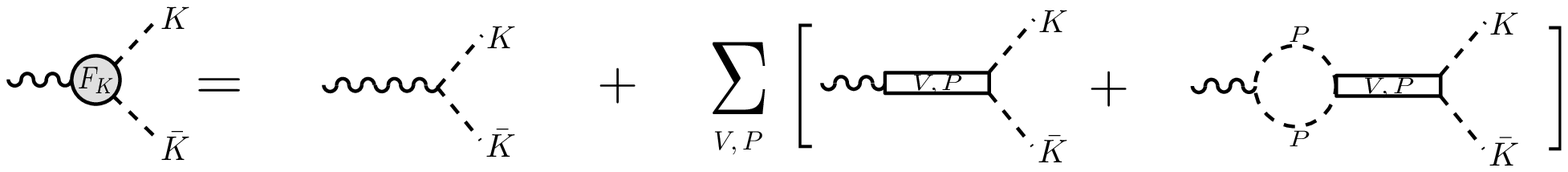}
\end{center}
\vspace{0.2cm}
\begin{center}
\includegraphics[width=110mm]{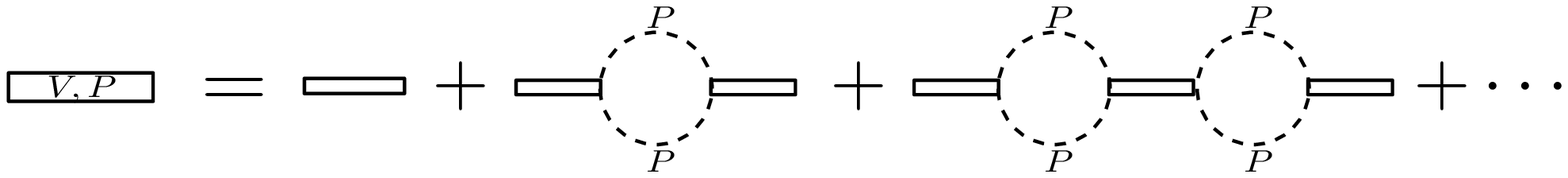}
\caption{Electromagnetic form factor of $K^+$ \ ($K^-$).}
    \label{fig:model}
\end{center}
\end{figure}

\begin{figure}[htbp]
\begin{center}
\includegraphics[width=140mm]{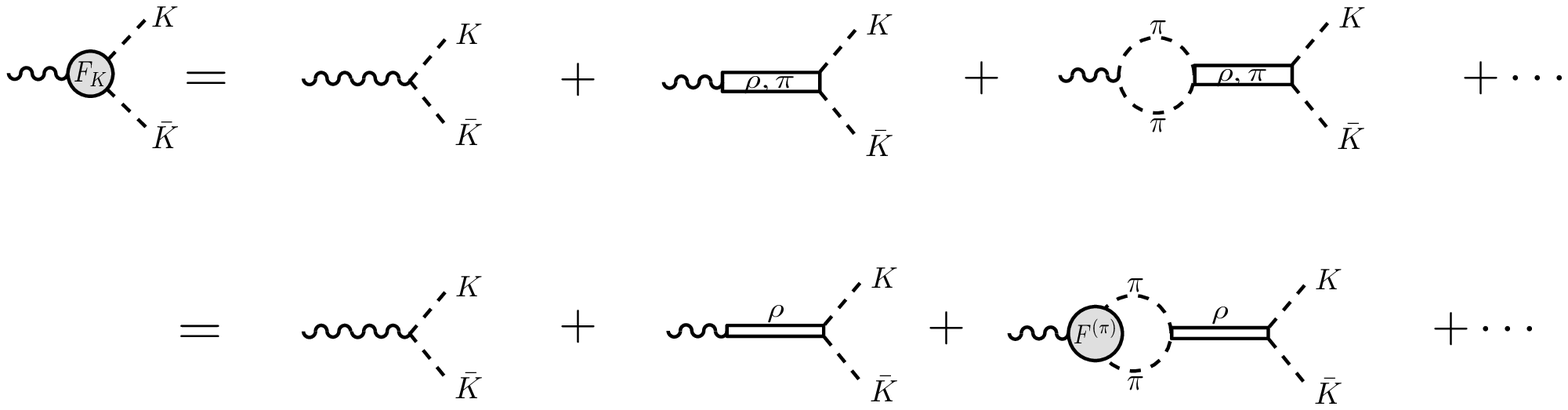}
\end{center}
\vspace{0.2cm}
\begin{center}
\includegraphics[width=70mm]{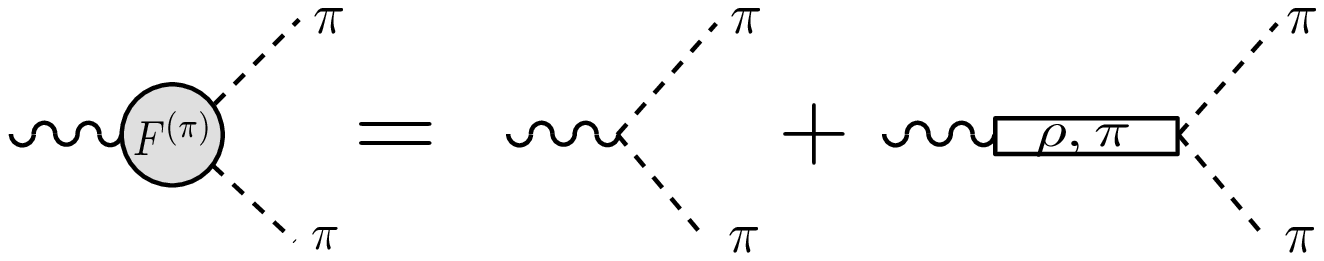}
\caption{Representation of kaon form factor in terms of pion
electromagnetic vertex.}
    \label{fig:somec}
\end{center}
\end{figure}

Fig.~\ref{fig:model} illustrates the model for the FF's including
self-energy and EM vertex loop corrections. The diagrams in
Fig.~\ref{fig:model} can also be re-arranged in a different way.
As an example, selecting in Fig.~\ref{fig:model} contribution from
$\rho\pi\pi$ vertex by taking $V=\rho$ and $P=\pi$ one obtains
Fig.~\ref{fig:somec}. This figure shows that the kaon FF contains
the pion EM vertex $F^{(\pi)}$ as a building block. For the
on-mass-shell pions $F^{(\pi)}$ would reproduce the pion FF.

The equations for the modified EM couplings read in terms of the
loop corrections
\begin{equation}
\frac{1}{f_V(s)} = \frac{1}{f_V^{(0)}} - \frac{\imath}{e\,s}
\sum_c {\rm\, Im} \Pi_{\gamma(c)V}(s),
\label{eq:E3}
\end{equation}
for $V =\rho^0, \omega, \phi$, and where index $c = (\pi^0\omega,
\pi\pi,$ $\pi^0\rho, KK,$ $3\pi, 3\pi-\rho\pi)$ stands for the
diagrams shown in Fig.~\ref{fig:loops_vertex}. The ``bare''
constants $f_V^{(0)}$ are real-valued, i.e. ${\rm Im} f_V^{(0)}
=0$. The modified couplings $f_V(s)$ at $s=m_V^2$ have to describe
the leptonic decay widths of the vector mesons
\begin{equation}
\bigl| f_V(s=m_V^2) \bigr|^2 = \frac{4}{3}\pi\alpha^2
\frac{m_V}{\Gamma(V \to e^+e^-)}.
\end{equation}
This allows us to find the bare couplings
\begin{equation}
\frac{1}{(f_V^{(0)})^2} =\frac{1}{\bigl| f_V (s=m_V^2) \bigr|^2}
\,-\, \frac{1}{e^2\,m_V^4}\Bigl( \sum_{c}\;{\rm Im}
\Pi_{\gamma(c)V}(s=m_V^2 ) \Bigr)^2 .
\end{equation}
Using the particle properties \cite{pdg} (see also
Table~\ref{tab:EM-couplings}) we obtain
 \beq f_\rho^{(0)} =5.0261,  \quad \quad  f_\omega^{(0)} =17.0601,
\quad \quad f_\phi^{(0)} = - 13.3824, \nonumber
 \eeq
and for arbitrary $s$ the real and imaginary parts of the
momentum-dependent couplings $f_V (s)$ are calculated from
(\ref{eq:E3}).

\begin{figure}[htbp]
\begin{tabular}{p{8.2cm}p{8.2cm}}
\includegraphics[width=80mm]{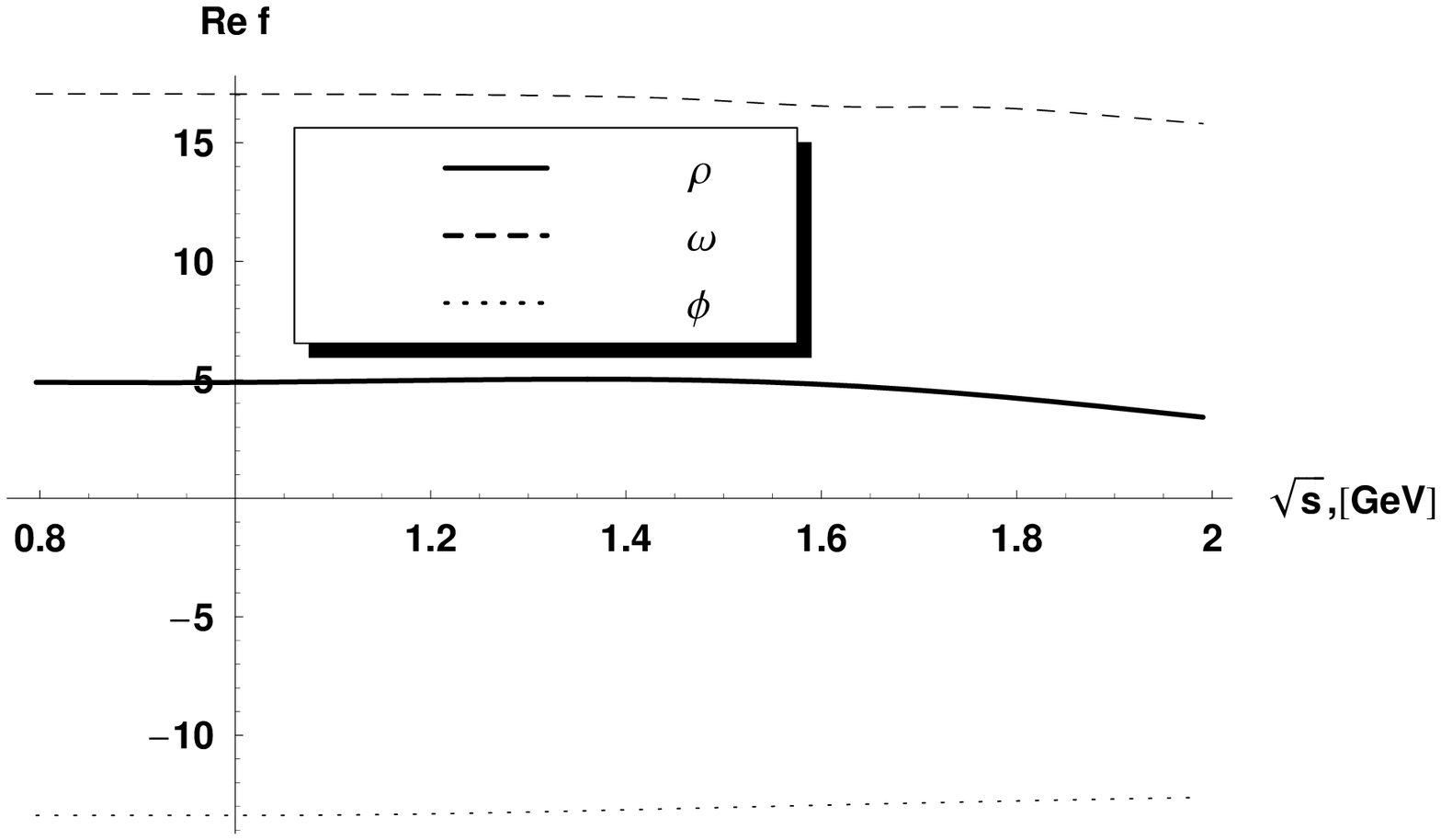}
&
\includegraphics[width=80mm]{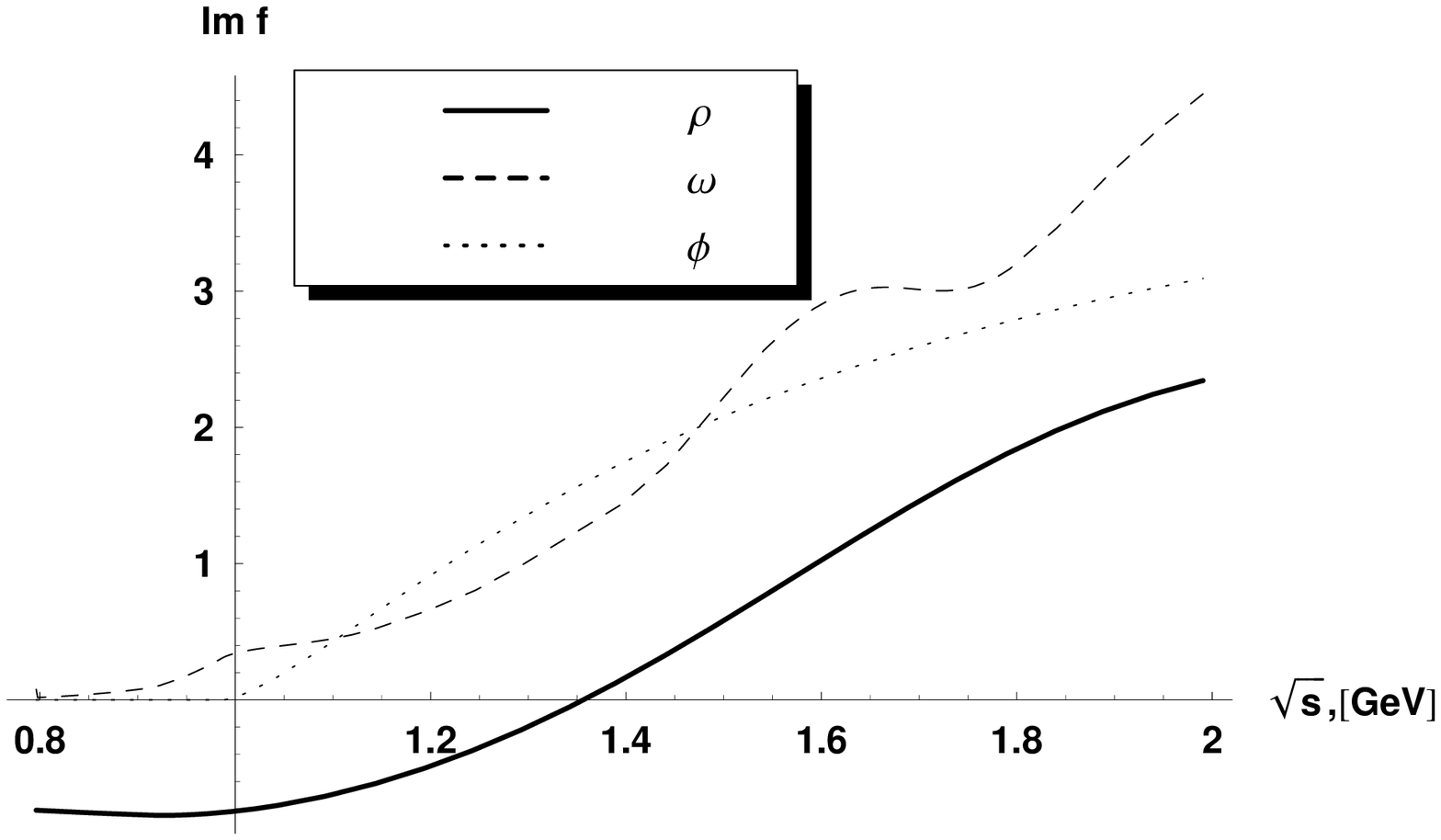}
\end{tabular}
\caption{Real (left) and imaginary (right) parts of the coupling
constants $f_V (s)$ for $\rho$, $\omega$ and $\phi$ mesons.}
\label{fig:EMcoupl}
\end{figure}

On the mass shells the imaginary parts  \ ${\rm Im\,}
\Pi_{\gamma(c)\omega}(s=m_\omega^2)$ and ${\rm Im\,}
\Pi_{\gamma(c)\phi}(s=m_\phi^2)$ turn out to be small and the
corresponding bare couplings $f_\omega^{(0)}$  and $f_\phi^{(0)}$
are very close to the values  ${\rm Re\,}f_\omega (s=m_\omega^2)$
and ${\rm Re\,}f_\phi (s=m_\phi^2)$. Nevertheless, off the mass
shells there remains certain $s-$dependence (see
Fig.~\ref{fig:EMcoupl}).

\subsection{Contribution from higher resonances}
\label{subsec:higher_resonances}

The contribution from the higher resonances $\rho',\omega',\phi'$ is
included by adding
\begin{equation}
\Delta F_{K^+} (s) = - \sum_{V' =\rho',
\omega',\phi'}\frac{g_{V'K^+ K^-}}{f_{V'}(s)}A_{V'}(s), \quad
\quad \Delta F_{K^0} (s) = -
\sum_{V'=\rho',\omega',\phi'}\frac{g_{V'K^0 \bar{K}^0}}{f_{V'}(s)}
A_{V'}(s), \label{eq:higher-res}
\end{equation}
to FF's in (\ref{eq:FFs}).

Though the current experimental data are not sufficient to
unambiguously find the couplings $f_{V'}$, \ $g_{V'K^+ K^-}$ and
$g_{V' K^0 \bar{K}^0}$, there is a way to constrain them. One can
use the branching ratios \cite{pdg}
\begin{subequations}
 \bea
\label{eq:data-rho'} \frac{\Gamma(\rho(1450)\to\pi^+\pi^-) \times
\Gamma(\rho(1450)\to e^+e^-)}{\Gamma_{tot}(\rho(1450))} &=&
\frac{1}{\Gamma_{tot}(\rho^\prime)} \frac{g_{\rho^\prime \pi\pi}^2
m_{\rho^\prime}}{3\cdot 16\pi} \bigl(1 - 4
\frac{m_\pi^2}{m_{\rho^\prime}^2} \bigr)^{3/2} \frac{e^4
m_{\rho^\prime}}{12 \pi f_{\rho^\prime}^2} \nonumber
\\
&=& 0.12 \;\text{keV}, \label{eq:data-rho'A}
\\
&=& \bigl(0.027\; \begin{matrix}+ 0.015 \\- 0.010 \end{matrix}\;
\bigr) \;\text{keV} \label{eq:data-rho'B}
 \eea
\end{subequations}
 \bea
\frac{\Gamma(\phi(1680)\to K^0_L K^0_S) \times
\Gamma(\phi(1680)\to e^+e^-)}{\Gamma_{tot}(\phi(1680))^2} &= &
\frac{1}{\Gamma_{tot}(\phi^\prime)^2} \frac{g_{\phi^\prime K^0
\bar{K}^0}^2 m_{\phi^\prime}}{3\cdot 16\pi} \bigl(1 - 4
\frac{m_K^2}{m_{\phi^\prime}^2} \bigr)^{3/2} \frac{e^4
m_{\phi^\prime}}{12 \pi f_{\phi^\prime}^2} \nonumber
\\
&=& (0.131 \pm 0.059) \times 10^{-6}
 \label{eq:data-phi'}
 \eea
and total widths  \ $\Gamma_{tot}(\rho^\prime) = 400 \pm 60$ MeV,
$\Gamma_{tot}(\omega^\prime) = 215 \pm 35$ MeV and
$\Gamma_{tot}(\phi^\prime) = 150 \pm 50$ MeV.

From these relations one finds ratios of the strong and EM
couplings, ${g_{\rho\prime K^+K^-}}/{f_{\rho\prime}}$ and
${g_{\phi\prime K^0 \bar{K}^0}}/{f_{\phi\prime}}$. We can also use
the $SU(3)$ relations for the {\it ratios} (see
Tables~\ref{tab:EM-couplings} and \ref{tab:strong-couplings}):
\begin{eqnarray}
 \frac{g_{\rho\prime K^+K^-}}{f_{\rho\prime}} :
\frac{g_{\omega\prime K^+K^-}}{f_{\omega\prime}} :
\frac{g_{\phi\prime K^+K^-}}{f_{\phi\prime}} &=& \frac{1}{2} :
\frac{1}{6} : \frac{1}{3}, \label{eq:SU3}  \\
\frac{g_{\rho\prime K^0\bar{K}^0}}{f_{\rho\prime}} :
\frac{g_{\omega\prime K^0\bar{K}^0}}{f_{\omega\prime}} :
\frac{g_{\phi\prime K^0\bar{K}^0}}{f_{\phi\prime}} &=&
-\frac{1}{2} : \frac{1}{6} : \frac{1}{3} \nonumber
\end{eqnarray}
and $|g_{\rho\prime K^+K^-}| = |g_{\rho\prime K^0\bar{K}^0}|$.

From (\ref{eq:data-rho'A}) and (\ref{eq:data-phi'}) we get the 2nd
column of Table~\ref{tab:higher-res}. It is seen that
${g_{\rho\prime K^+K^-}}/{f_{\rho\prime}}$ and ${g_{\phi\prime
K^+K^-}}/{f_{\phi\prime}}$ approximately follow $SU(3)$ symmetry
(\ref{eq:SU3}) that allows us to obtain parameters for
$\omega^{\prime}-$meson. Note that $SU(3)$ symmetry is not
satisfied for parameters calculated from (\ref{eq:data-rho'B}) and
(\ref{eq:data-phi'}), which are shown in the 3rd column of
Table~\ref{tab:higher-res}. Above estimates for parameters of the
``primed'' resonances are used further in calculations. For
comparison in the 4th and 5th columns we present the values
obtained from the ``constrained'' and ``unconstrained'' fits to
data in Ref.~\cite{Bruch_04}.

\begin{table}
\begin{center}
\begin{tabular}{|c|c|c|c|c|}
\hline Ratios    & eqs.(\ref{eq:data-rho'A}), (\ref{eq:data-phi'})
& eqs.(\ref{eq:data-rho'B}), (\ref{eq:data-phi'}) &
constrained fit \cite{Bruch_04} & unconstrained fit \cite{Bruch_04} \\
\hline
 $\frac{g_{\rho' K^+ K^-}}{f_{\rho'}}$   & $-0.063 \pm 0.005$ & $-0.03 \pm 0.0024 \pm 0.007$ &
 $-\frac{0.112}{1.195}\frac{g_{\rho K^+K^-}}{f_{\rho}}= -0.056$    &
$-\frac{0.124}{1.139}\frac{g_{\rho K^+K^- }}{f_{\rho}}=-0.065 $  \\
\hline $\frac{g_{\omega' K^+ K^-}}{f_{\omega'}}$ & $
\frac{g_{\rho' K^+ K^-}}{3 f_{\rho'}}$ & $\frac{g_{\rho' K^+
K^-}}{3 f_{\rho'}}$ & $-\frac{0.112}{1.195}\frac{g_{\omega
K^+K^-}}{f_{\omega}}=-0.016$ &
$-\frac{0.018}{1.467}\frac{g_{\omega K^+K^- }}{f_{\omega}}= -0.002$ \\
\hline $\frac{g_{\phi' K^+ K^-}}{f_{\phi'}}$ & $-0.036 \pm
0.012\pm 0.008$ & $-0.036 \pm 0.012 \pm 0.008$  &
$-\frac{0.018}{1.018}\frac{g_{\phi K^+K^-}}{f_{\phi}}= -0.005$ &
$-\frac{0.001}{0.999}\frac{g_{\phi K^+K^-}}{f_{\phi}} \approx 0.0$
 \\
 \hline
\end{tabular}
\end{center}
\caption{Ratios of coupling constants for resonances $\rho',
\omega', \phi'$ (only couplings to $K^+ K^-$ are shown for
brevity). Errors are due to uncertainty in decay widths (first)
and branching ratios (second). The signs of couplings are chosen
opposite to those for $\rho, \omega, \phi$.}
\label{tab:higher-res}
\end{table}

There still remains a sign ambiguity in these ratios. One may
determine the relative sign of the $\rho', \omega', \phi'$
contribution with respect to the $\rho, \omega, \phi$ one using
the following reasoning. On the basis of quark counting rule in
perturbative QCD  \cite{Lepage_1980} the FF behaves like  $
F_{K^+}(s) \to a/s$ \ at $s \to -\infty$, where $a = -16 \pi
F_{\pi}^2 \alpha_s (s)$. We obtain in this limit from
(\ref{eq:FFs}) and (\ref{eq:higher-res})
 \bea
&& F_{K^+}(s)  \xrightarrow[s \to -\infty]{} b + \frac{a}{s},
\label{eq:large-s-limit}
\\
&& b =  1 - \sum_{V=\rho,\omega,\phi}\frac{g_{VK^+K^-}}{f_V}\; -
\sum_{V^\prime=\rho',\omega',\phi'}\frac{g_{V^\prime
K^+K^-}}{f_{V^\prime}},  \label{eq:constant_b} \\
&& a = - \sum_{V=\rho,\omega,\phi}\frac{g_{VK^+K^-}m_V^2}{f_V}\; -
\sum_{V^\prime=\rho',\omega',\phi'}\frac{g_{V^\prime
K^+K^-}m_{V'}^2}{f_{V^\prime}}. \label{eq:constant_a}
\end{eqnarray}
For the correct asymptotic behavior the constant $b$ in
(\ref{eq:constant_b}) should be equal to zero. One can check
however that the contribution from the lowest resonances $\rho,
\omega, \phi$ with the couplings $g_{VK^+K^-}/f_V$ taken from
experiment does not lead to vanishing of $b$. If we add the
contribution from the higher resonances $\rho', \omega', \phi'$
choosing the negative relative sign of couplings $g_{V'
K^+K^-}/f_{V'}$ with respect to $g_{VK^+K^-}/f_V$, then the
constant $b$ becomes closer to zero. In this way the asymptotic
behavior of the form factors is partially improved. The sign
``minus'' for the higher resonances also improves significantly
description of the experimental form factors which are calculated
in sect.~\ref{sec:Results}. Note also that the constant $a$ in
(\ref{eq:constant_a}) takes negative values with the parameters in
Table~\ref{tab:higher-res}.

There is also experimental information on the relative phases of
the higher resonance contribution in a different reaction $e^+e^-
\to \pi^+\pi^-\pi^0$ \cite{Achasov0305049,Achasov_PL462}. Our
choice of the relative sign is in line with the findings of these
papers.

\section{Results of calculation}
\label{sec:Results}

\subsection{Kaon form factors and cross section of $e^+e^- \to K\bar{K}$ reaction}
\label{subsec:Results_kaon}

\begin{figure}[htbp]
\includegraphics[width=82mm]{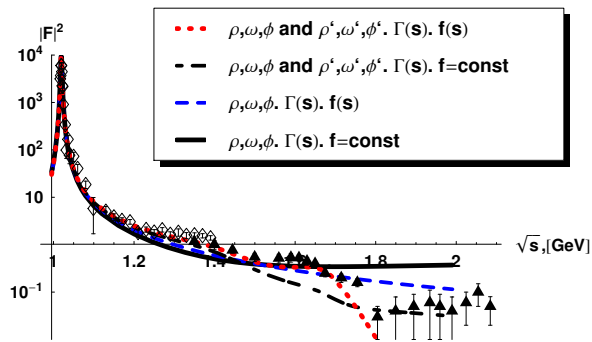}
    \caption{Charged kaon EM form factor in the time-like region. Data:
diamonds (open) are from \cite{expIvanov}, triangles -- from
\cite{expBisello}.}
    \label{fig:ff_charged_1}
\end{figure}

\begin{figure}[htbp]
\includegraphics[width=82mm]{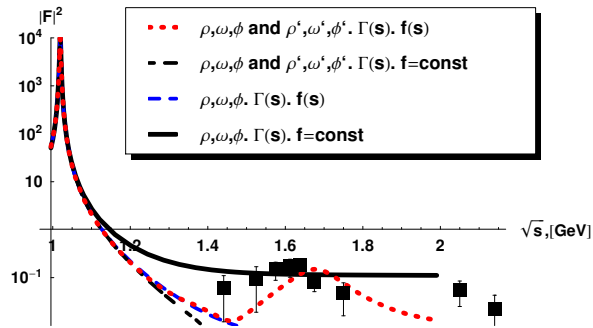}
    \caption{Neutral kaon EM form factor in the time-like region. Data (boxes) are from
\cite{expMane}.}
    \label{fig:ff_neutral_1}
\end{figure}

\begin{figure}[htbp]
\includegraphics[width=80mm]{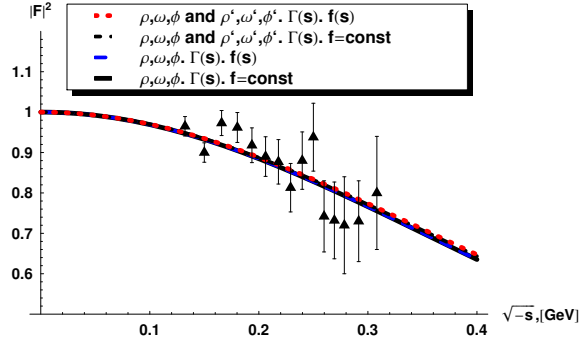}
\caption{Charged kaon EM form factor in the space-like region.
Data are from \cite{expAmendolia}.}
    \label{fig:kaoncharged_1_ff}
\end{figure}

In this section we present results for the charged and neutral
kaons. The FF's are calculated from (\ref{eq:A1}), (\ref{eq:A2}),
(\ref{eq:higher-res}), and the $e^+ e^-$ annihilation cross
section from (\ref{eq:cross-section}).
Figs.~\ref{fig:ff_charged_1} and \ref{fig:ff_neutral_1} show FF in
the time-like region, while Fig.~\ref{fig:kaoncharged_1_ff} shows
FF in the space-like region.

\begin{figure}[htbp]
    \centering
\begin{tabular}{cc}
    \includegraphics[width=75mm]{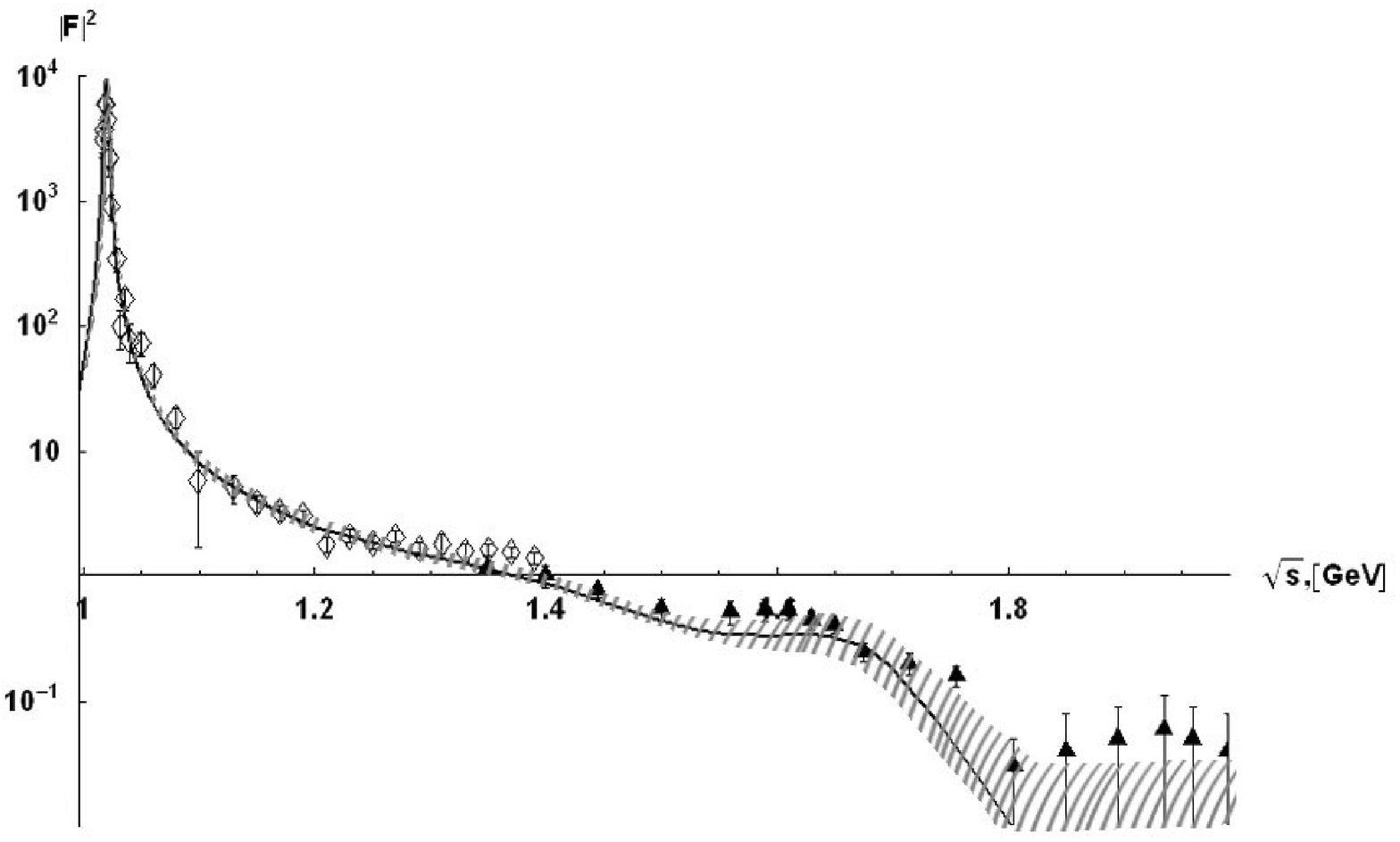}
    &
    \includegraphics[width=75mm]{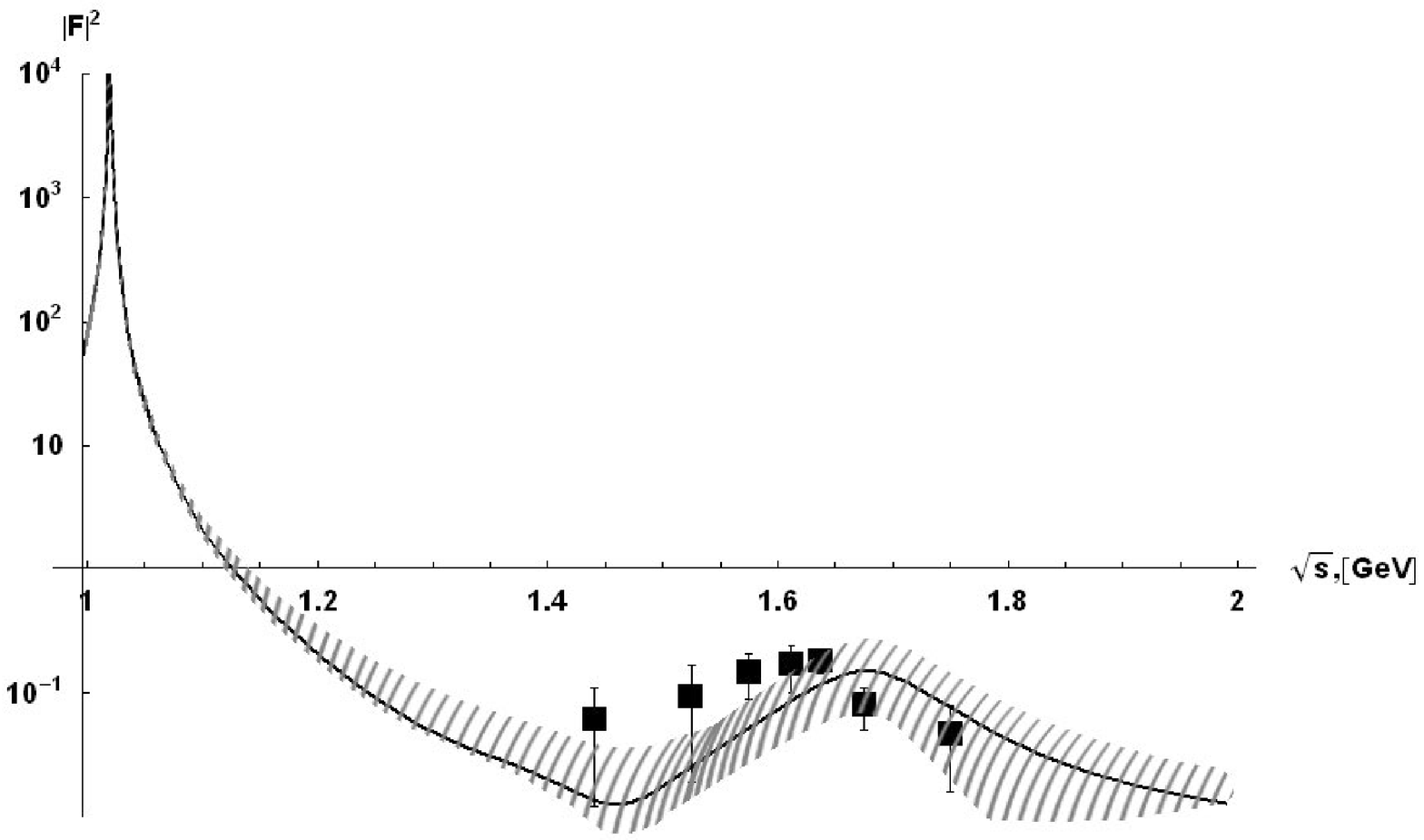}
\end{tabular}
  \caption{Form factors of charged kaon (left) and neutral kaon (right).
  Hatched area shows error corridor caused by uncertainties
  in parameters of vector mesons. Data are the same
  as in Figs.~\ref{fig:ff_charged_1} and \ref{fig:ff_neutral_1}. }
  \label{fig:korid}
\end{figure}

\begin{figure}[htbp]
\includegraphics[width=82mm]{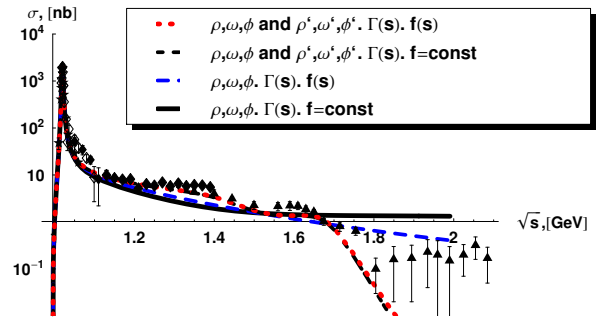}
    \caption{Total cross section of $e^+e^-$ annihilation into charged kaons.
Data: stars are from \cite{expAchasov}, diamonds (filled) -- from
\cite{expDolinsky}, triangles -- from \cite{expBisello}, diamonds
(open) -- from \cite{expIvanov}.}
    \label{fig:cs_charged_1}
\end{figure}
\begin{figure}[htbp]
\includegraphics[width=82mm]{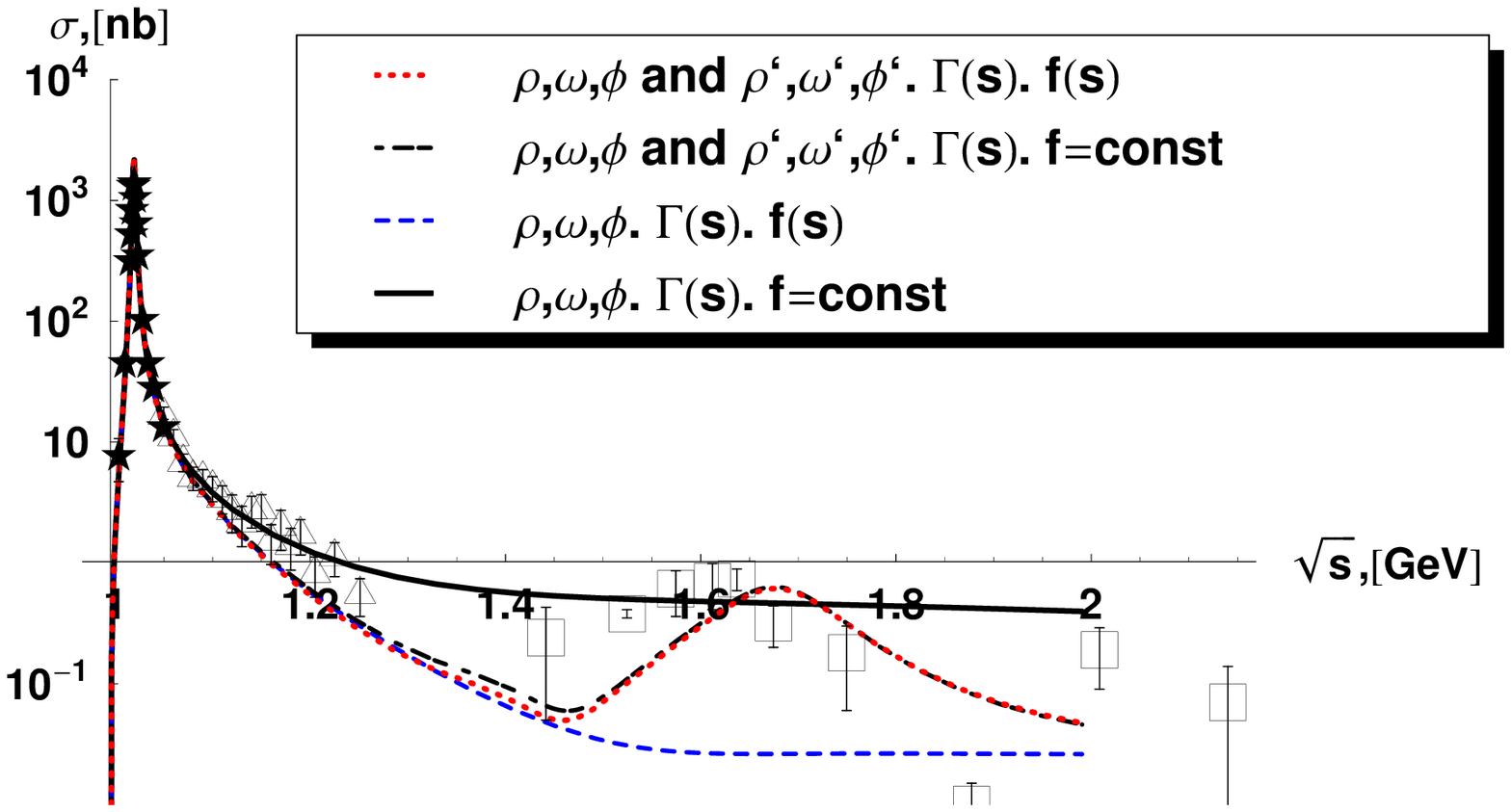}
    \caption{Total cross section of $e^+e^-$ annihilation into neutral kaons.
Data: stars are from \cite{expAchasov}, triangles (open) -- from
\cite{expAkhmetshin}, boxes (open) -- from \cite{expMane}.}
    \label{fig:cs_neutral_1}
\end{figure}

The solid curves (see legends in the plots) represent a simple
VMD-like model in which only $\rho$, $\omega$ and $\phi$
resonances are included. The meson widths are taken $s$-dependent
(see sect.~\ref{sec:self-energy}) while the couplings of vector
mesons to photon $f_V$ are independent of momentum. As known, such
a model can describe experiment only in vicinity of the
$\phi(1020)$ resonance.

The long-dashed curves include in addition the momentum-dependent
EM couplings. It is seen that this inclusion does not improve
considerably the description. These results suggest the need for
further complication of the model.

Taking into consideration the excited resonances $\rho ^\prime$,
$\omega ^\prime$ and $\phi ^\prime$ with momentum-dependent widths
and constant couplings $f_V$ we obtain the dot-dashed curves. The
corresponding calculation improves the agreement with available
data in comparison with the simple VMD model. The contribution
from $\rho(1450)$, $\omega(1420)$ and $\phi(1680)$ resonances is
noticeable.

The short-dashed curves represent the most advanced calculation.
These curves include the momentum-dependent widths for all
intermediate states, ``dressed'' EM vertices (for $\rho$, $\omega$
and $\phi$ resonances) and cut-off FF's in the self-energies and
EM vertices (see (\ref{eq:K1}) and (\ref{eq:K2})). We have not
attempted to develop vertex ``dressing'' for the higher resonances
due to the present experimental uncertainties in their decay
rates. We should note that the authors of Ref.~\cite{Bruch_04}
also obtained a good description of the data by fixing the values
of the parameters $f_V$ from the fit. In our procedure of
``dressing'' the couplings a reasonable agreement is achieved
without fitting the parameters.

It is of interest to study sensitivity of calculated FF's to
uncertainties in the model parameters. The uncertainties in
parameters $f_V$ and $g_{V^\prime KK}/ f_{V^\prime}$ for vector
mesons (see Tables~\ref{tab:EM-couplings} and
\ref{tab:higher-res}) arises due to a limited experimental
accuracy in decay widths and branching ratios, especially for the
primed resonances. Fig.~\ref{fig:korid} demonstrates how these
model uncertainties influence FF's.

Figs.~\ref{fig:cs_charged_1} and \ref{fig:cs_neutral_1} represent
the $e^+ e^- \to K\bar{K}$ cross section. They are compared with
all available data. Finally the plot in
Fig.~\ref{fig:kaoncharged_1_ff} shows the charged kaon FF in the
space-like region of photon momentum. This figure demonstrates
good agreement with data \cite{expAmendolia}. At the same time,
the FF in the region of $-q^2 < 0.16$ GeV$^2$ is not sensitive to
ingredients of the model.

Note that kaon FF's in the space-like region have been studied
earlier in various approaches: quark Dyson-Schwinger equation
\cite{Maris}, ChPT in order ${\cal O}(p^{6})$ \cite{Bijnens},
quark linear $\sigma-$model \cite{Scadron}, model based on pion FF
and data on $K^+ \to \pi^+ e^+ e^-$ \cite{Lowe}, relativistic
constituent quark model \cite{Cardarelli}.

From Figs.~\ref{fig:ff_charged_1} - \ref{fig:cs_neutral_1} one can
conclude that the developed model is consistent with the
experimental information \cite{expAkhmetshin},
\cite{expIvanov,expBisello,expMane,expAchasov,expDolinsky} in the
1--1.75 GeV energy region. Although the most important
contributions are included in the model, some deviations show up
in the higher-energy region. They may be attributed to other
intermediate states or experimental uncertainty in the parameters
of $\rho ^\prime$, $\omega ^\prime$ and $\phi ^\prime$ resonances.

\subsection{Contribution of $K^+ K^-$ and $K^0 \bar{K}^0$ channels
to anomalous magnetic moment of muon}
\label{subsec:Results_muon}

The contribution of the $K \bar{K}-$channel to $a_\mu$ is directly
related to $F_{K^+ }(s)$ and $F_{K^0}(s)$ via the dispersion
integral \cite{Brodsky_68} (see also \cite{Davier_2003}):
\begin{eqnarray}
a^{had,K\bar{K}}_{\mu}&=& \frac{\alpha^2}{3
\pi^2}\int_{4m^2_K}^{\infty} W(s)R(s) \frac{\mathrm{d}s}{s},
\label{eq:a-mu} \\
W(s)&=& \int_0^1 \frac{x^2 (1-x)}{x^2+(1-x) s/m^2_{\mu}}
 \mathrm{d}x   \nonumber \\
& =& y^2(1-\frac{y^2}{2}) +(1+y)^2
(1+\frac{1}{y^2})[\ln(1+y)-y+\frac{y^2}{2}]
 +\frac{1+y}{1-y} y^2 \ln y, \nonumber
\end{eqnarray}
where $y=(1-\beta)/(1+\beta)$, \ $\beta =(1-4m_\mu^2/s)^{1/2}$,
$m_{\mu}$ is the muon mass, and $R(s)$ is the ratio of the cross
sections
 \begin{equation}
 R(s) =\frac{\sigma(e^+e^- \to K\bar{K})}{\sigma(e^+e^- \to
\mu^+ \mu^-)}=
\frac{(1-\frac{4m_{K}^2}{s})^{3/2}}{4(1+2\frac{m_\mu^2}{s})(1-\frac{4m_{\mu}^2}{s})^{1/2}
} |F_{K }(s)|^2. \label{eq:CS_ratio}
\end{equation}
Here $\sigma(e^+e^- \to K\bar{K})$ is the `bare' (lowest order)
cross section, not including initial-state radiation, $\gamma e^+
e^-$ vertex corrections and vacuum polarization effects.

\begin{table}
\begin{center}
\begin{tabular}{|c|c|c|c|}
\hline
 & $K^+ K^-$  & $K^0 \bar{K}^0$ & total $K \bar{K}$ \\
\hline calculation based on
(\ref{eq:data-rho'A}),~(\ref{eq:data-phi'})& $19.18 \pm 0.45$ &
$15.60 \pm 0.40$ & $34.78 \pm 0.85$
\\
\hline calculation based on
(\ref{eq:data-rho'B}),~(\ref{eq:data-phi'}) & $19.00 \pm 0.51$  &
$15.70 \pm 0.39$ & $34.70 \pm 0.90$
\\
\hline average & $19.06 \pm 0.57$  & $15.64 \pm 0.44$ & $34.70 \pm
1.01$
\\
\hline
\end{tabular}
\end{center}
\caption{Contribution of $K\bar{K}-$channels to anomalous magnetic
moment of the muon  $a_\mu^{had, K \bar{K}}$ in units $10^{-10}$.}
\label{tab:AMM}
\end{table}

The calculated values are presented in Table~\ref{tab:AMM}
together with the inaccuracy caused by uncertainty in the
parameters of the model. The $K\bar{K}-$channels contribute
$(34.70 \pm 1.01) \times 10^{-10}$ to the hadronic contribution
$a_\mu^{had}$. The latter, according to Ref.~\cite{Davier_2003},
is $a_\mu^{had,LO} = (696.3 \pm 6.2_{exp} \pm 3.6_{rad}) \times
10^{-10}$ (in LO). We can compare our result with that of
\cite{Davier_2003} for the $K\bar{K}-$channels (see Table~1 in
\cite{Davier_2003}, entries for $\phi$ and $K \bar{K}$
contributions in $e^+ e^-$ annihilation):
\begin{eqnarray} && (35.71
\pm 0.84_{exp} \pm 0.20_{rad})_{\phi} \times 83.1\% + (4.63 \pm
0.40_{exp} \pm 0.06_{rad})_{K^+ K^-} \nonumber \\
 && + (0.94 \pm 0.10_{exp} \pm 0.01_{rad})_{K_S K_L}
 = 35.24 \pm  1.20_{exp} \pm 0.24_{rad}
 \nonumber
\end{eqnarray}
in units $10^{-10}$. It is seen that our model gives $a_\mu^{had,
K \bar{K}}$ very close to the value obtained from $e^+ e^- $
annihilation cross sections.

\section{Conclusions}
\label{sec:Conclusions}

We developed a model for electromagnetic FF's of the charged and
neutral kaon in the time-like ($\sqrt{s} = $1--2 GeV) and
space-like ($s \le 0$) regions of the photon momentum. The model
is based on the chiral perturbation theory (ChPT) with vector
mesons \cite{Ecker1}. The photon vector-meson vertices by
construction depend on the photon momentum $q$ and are explicitly
gauge invariant. This ensures the correct normalization of FF's at
the real-photon point $ F_{K^+ }(q^2 =0)=1$ and $F_{K^0}(q^2 =0)
=0$ without need for parameters adjustment.

Beyond the tree level the model includes certain loop corrections,
such as self-energy polarization operators in vector-meson
propagators, and ``dressed'' photon-meson vertices. For
construction of the vertices we apply chiral Lagrangians
\cite{Ecker1,Rudaz1,Prades} and WZW Lagrangians \cite{WZW1,WZW2}.
The coupling constants are fixed from experimental decay widths of
resonances. The parameters in the odd-intrinsic-parity sector are
also compared with those obtained from a generalization of the WZW
approach for vector and axial-vector mesons \cite{Rudaz1} and
extended Nambu-Jona-Lasinio model \cite{Prades}.

Comparison of the calculations with available data for the FF's
and $e^+ e^-$ annihilation cross sections indicates that the model
is consistent with experimental information in the
$\sqrt{s}$=1--1.75 GeV energy region. Using the dressed
photon-meson vertices and adding the resonances
$\rho^\prime=\rho(1450)$, $\omega^\prime=\omega(1420)$ and
$\phi^\prime =\phi(1680)$ considerably improve description of the
data. A reasonable agreement is achieved without fitting the
parameters of the model.

Although the most important contributions are included, deviations
from the data appear at high energies, $\sqrt{s} \sim 2$ GeV.
Those may be attributed to missing contribution from the higher
(``double-primed'') resonances $\rho(1700)$ and $\omega(1650)$
\cite{pdg}, or possibly experimental uncertainty in parameters of
the $\rho^\prime$, $\omega^\prime$ and $\phi^\prime$.

The charged kaon FF extended to the space-like region agrees with
the data \cite{expAmendolia} at relatively small values of
momentum transfer, $-q^2 < 0.16$ GeV$^2$, though all variants of
the model lead to the close results. Current measurements at JLab
\cite{JLab-E98} of the longitudinal cross section in the reactions
$ep \to e\Lambda K^+$ and $ep \to e \Sigma^0 K^+$  will provide
information on kaon FF at momentum transfer up to $-q^2 \sim 3$
GeV$^2$. These experiments can help to discriminate between
variants of the present model, as well as between various
theoretical approaches
\cite{Maris,Bijnens,Scadron,Lowe,Cardarelli}.

The calculated FF's allowed us also to evaluate leading-order
contribution of the $K \bar{K}-$channel to the muon anomalous
magnetic moment $a_\mu$. The calculated value $a_\mu^{had, K^+ K^-
+ K^0 \bar{K}^0} = (34.70 \pm 1.01) \times 10^{-10}$ is in
agreement with result \cite{Davier_2003} obtained from the
experimental $e^+ e^- $ annihilation cross sections.

\section*{Acknowledgement}
We would like to thank S.~Eidelman and N.~Merenkov for useful
suggestions and remarks. This work is supported by grant INTAS
Ref. Nr 05-1000008-8328.

\appendix{}
\section{Chiral Lagrangian for pseudoscalar, vector mesons and photon}
\label{App:A}

\setcounter{equation}{0}
\def\theequation{A.\arabic{equation}}

\hspace{0.5cm} In the even intrinsic-parity sector we choose
$\mathcal{O}(p^{2})$ chiral Lagrangian derived by Ecker et al.
\cite{Ecker1}, which includes interaction of pseudoscalar,
polar-vector, axial--vector mesons and photon. We omit in this
Lagrangian contribution from axial-vector mesons which is not
relevant here. The explicit form is then
\begin{eqnarray}
L &=&\frac{F_{\pi }^{2}}{4}\mathrm{Tr} (D_{\mu }UD^{\mu
}U^{\dagger} +\chi U^{\dagger} +\chi^{\dagger} U)
-\frac{1}{4}F_{\mu \nu} F^{\mu \nu}
\nonumber \\
&& - \frac{1}{2} \mathrm{Tr}(\nabla^\lambda V_{\lambda \mu}
\nabla_\nu V^{\nu \mu} - \frac{1}{2}M_V^2 \, V_{\mu \nu} V^{\mu
\nu} )  \nonumber \\
&& +\frac{F_{V}}{2\sqrt{2}}\mathrm{Tr}(V_{\mu \nu }f_{+}^{\mu \nu
})+\frac{ iG_{V}}{\sqrt{2}}\mathrm{Tr}(V_{\mu \nu }u^{\mu }u^{\nu
}),  \label{eq:B1}
\end{eqnarray}
where $U=\exp (i\sqrt{2}\Phi /F_{\pi })$, \ $\Phi $ describes the
pseudoscalar mesons, $V_{\mu \nu }$ is antisymmetric field
describing the vector mesons, and $F^{\mu \nu }=\partial ^{\mu
}B^{\nu }-\partial ^{\nu }B^{\mu }$ is the EM tensor for the gauge
photon field $B^\mu$. The pion weak-decay constant $F_\pi$ and
coupling constants $F_{V},G_{V}$ are specified in
sect.~\ref{sec:Formalism}.

The covariant derivative $D_\mu$ is defined as
\begin{equation}
D_\mu U = \partial_\mu U + \imath e B_\mu[U,Q]
\end{equation}
with quark charge matrix $Q =
\mathrm{diag}(\frac{2}{3},-\frac{1}{3},-\frac{1}{3})=\frac{1}{2}\lambda_3
+ \frac{1}{2\sqrt{3}}\lambda_8$, $ \ \lambda_a$ ($a=1,...,8$) are
the Gell-Mann matrices and $\lambda_0 = \sqrt{\frac{2}{3}}\, {1}$.
Also in (\ref{eq:B1}) $\chi$ is proportional to quark mass matrix
${\cal M} =\mathrm{diag}(m_u, m_d, m_s )$ and \bea && U = u^2,
\quad \quad \quad u^\mu = \imath u^+ (D^\mu U)u^+ ,
\nonumber \\
&& f_\pm{}^{\mu\nu} =  e F^{\mu \nu} (u Q u^+ \pm u^+ Q u). \eea
For other definitions see \cite{Ecker1}.

In the odd intrinsic-parity sector, Lagrangian
\cite{Ecker3,Prades} in the vector formulation for vector mesons
is applied. The pseudoscalar-vector interaction of chiral order
${\cal O}(p^3)$ and ${\cal O}(p^2)$ is
\begin{equation}
L_{int} = i \theta_V \epsilon_{\mu \nu \alpha \beta} \mathrm{Tr}
(V^\mu u^\nu u^\alpha u^\beta ) + h_V \epsilon_{\mu \nu \alpha
\beta} \mathrm{Tr} (V^\mu \{ u^\nu, f^{\alpha \beta}_{+} \} ) +
\sigma_V  \epsilon_{\mu \nu \alpha \beta} \mathrm{Tr} (V^\mu \{
u^\nu, \partial^{\alpha} V^{\beta } \} ), \label{eq:odd-parity}
\end{equation}
where $\{a,b\}\equiv ab+ba$ and we omitted piece with axial-vector
fields, higher-order terms in vector fields $V^\mu$, and some
other terms which give no contribution to form factors in the
present model. The coupling parameters $\theta_V$, $h_V$ and
$\sigma_V$ are discussed in sect.~\ref{sec:Formalism}.

The $SU(3)$ flavor symmetry of the meson multiplets is
conventionally accounted for by using notation for
pseudoscalar-meson ($J^P=0^-$) octet, and for the vector-meson
($J^{PC}=1^{--}$) nonet in vector and tensor formulation,
respectively
\begin{eqnarray}
\Phi &=& \frac{1}{\sqrt{2}} \sum_{a=1}^8 \lambda_a \phi^a, \quad
\quad  \quad  V_\mu = \frac{1}{\sqrt{2}} \sum_{a=1}^8 \lambda_a
V^a_{\mu} +\frac{1}{\sqrt{2}}\lambda_0 V^0_{\mu}, \nonumber \\
V_{\mu \nu} &=& \frac{1}{\sqrt{2}} \sum_{a=1}^8 \lambda_a V^a_{\mu
\nu} +\frac{1}{\sqrt{2}}\lambda_0 V^0_{\mu \nu}.
\label{eq:octet-nonet}
\end{eqnarray}

In the present paper we do not need $K^{*\pm}$, $K^{*0}$,
$\bar{K^{*0}}$ and $\eta$ mesons and it is sufficient to take
\begin{eqnarray}
\Phi &=& \frac{1}{\sqrt{2}} (\pi_1 \lambda_1 + \pi_2 \lambda_2 +
\pi_3 \lambda_3 + K_1 \lambda_4 + K_2 \lambda_5 + K_3 \lambda_6 +
K_4 \lambda_7),
\\
V^\mu &=& \frac{1}{\sqrt{2}} (\rho_1^\mu \lambda_1 + \rho_2^\mu
\lambda_2 + \rho_3^\mu \lambda_3 + \omega^\mu (\lambda_0
\sin\theta + \lambda_8 \cos\theta) + \phi^\mu (\lambda_0
\cos\theta - \lambda_8\sin\theta) ),
\\
V^{\mu\nu}&=& \frac{1}{\sqrt{2}} (\rho_1^{\mu\nu} \lambda_1 +
\rho_2^{\mu\nu} \lambda_2 + \rho_3^{\mu\nu} \lambda_3 +
\omega^{\mu\nu} (\lambda_0 \sin\theta + \lambda_8 \cos\theta) +
\phi^{\mu\nu} (\lambda_0 \cos\theta - \lambda_8\sin\theta) ),
\end{eqnarray}
where the  physical fields are defined as
\begin{align*}
\pi^\pm = \frac{1}{\sqrt{2}} (\pi_1 \mp \imath \pi_2), && K^\pm =
\frac{1}{\sqrt{2}} (K_1 \mp \imath K_2),
\\
K^0 = \frac{1}{\sqrt{2}} (K_3 - \imath K_4), && \bar{K^0} =
\frac{1}{\sqrt{2}} (K_3 + \imath K_4),
\\
\rho^\pm = \frac{1}{\sqrt{2}} (\rho_1 \mp \imath \rho_2), &&
\rho^0 = \rho_3, \;\; \; \; \pi^0 = \pi_3 .
\end{align*}

Mixing angle $\theta$ can be estimated from experiment. It is
convenient to use $\omega-\phi$ mixing parameter \cite{Klingl}
$\epsilon_{\omega\phi} = {\sin}(\theta_{0} - \theta) = 0.058$
which naturally appears for OZI-forbidden processes. The ``ideal
mixing'' (with ${\sin}\theta_{0} = \sqrt{{2}/{3}}$,
$\;{\cos}\theta_{0} = 1 / \sqrt{3}$) \ corresponds to quark
content of the vector mesons: $\omega =
(\bar{u}u+\bar{d}d)/\sqrt{2}$ and $\phi = \bar{s}s$.

\end{document}